\documentclass[12pt,preprint]{aastex}
\usepackage{emulateapj5}
 
\newcommand{\cf}{{\ifmmode{C_f}\else{$C_{f}$}\fi}}
\newcommand{\zem}{{\ifmmode{z_{em}}\else{$z_{em}$}\fi}}
\newcommand{\zabs}{{\ifmmode{z_{abs}}\else{$z_{abs}$}\fi}}
\newcommand{\kms}{{\ifmmode{{\rm km~s}^{-1}}\else{km~s$^{-1}$}\fi}}
\newcommand{\delv}{{\ifmmode{\Delta v}\else{$\Delta v$}\fi}}
\newcommand{\voff}{{\ifmmode{v_{shift}}\else{$v_{shift}$}\fi}}
\newcommand{\cmm}{{\ifmmode{{\rm cm}^{-2}}\else{cm$^{-2}$}\fi}}
\newcommand{\cmmm}{{\ifmmode{{\rm cm}^{-3}}\else{cm$^{-3}$}\fi}}
\newcommand{\nhi}{{\ifmmode{N_{\rm H\;I}}\else{$N_{\rm H\;I}$}\fi}}
\newcommand{\nhe}{{\ifmmode{N_{\rm He\;II}}\else{$N_{\rm He\;II}$}\fi}}
\newcommand{\logn}{{\ifmmode{\log\; N}\else{$\log\; N$}\fi}}
\newcommand{\lognciv}{{\ifmmode{\log\; N_{\rm C\; IV}}\else{$\log\; N_{\rm C\; IV}$}\fi}}

\def\lsim{\lower0.3em\hbox{$\,\buildrel <\over\sim\,$}}
\def\gsim{\lower0.3em\hbox{$\,\buildrel >\over\sim\,$}}

\newcounter{species} 
\def\ion#1#2{\setcounter{species}{#2}#1$\;${\scriptsize\Roman{species}}\relax}

\newcommand{\lya}{Ly$\alpha$}

\slugcomment{\today}
\shorttitle{Dramatically Variable Mini-BAL in HS~1603+3820}
\shortauthors{Misawa et al.}

\begin{document}

\title{Results of Monitoring the Dramatically Variable C~IV Mini-BAL
   System in the Quasar HS~1603+3820\altaffilmark{1}}

\altaffiltext{1}{Based on data collected at Subaru Telescope, which is
operated by the National Astronomical Observatory of Japan.}

\author{Toru Misawa\altaffilmark{2}, 
        Michael Eracleous\altaffilmark{2,3,4}, 
        Jane C. Charlton\altaffilmark{2}, and
        Nobunari Kashikawa\altaffilmark{5,6}}

\altaffiltext{2}{Department of Astronomy \& Astrophysics, The
  Pennsylvania State University, University Park, PA 16802}
\altaffiltext{3}{Department of Physics \& Astronomy, Northwestern
  University, 2131 Tech Drive, Evanston, IL 60208}
\altaffiltext{4}{Center for Gravitational Wave Physics, The
  Pennsylvania State University}
\altaffiltext{5}{National Astronomical Observatory, 2-21-1 Osawa,
  Mitaka, Tokyo 181-8588, Japan}
\altaffiltext{6}{Department of Astronomical Science, Graduate
  University for Advanced Studies, 2-21-1 Osawa, Mitaka, Tokyo
  181-8588, Japan}

\email{misawa, mce, charlton@astro.psu.edu, kashik@optik.mtk.nao.ac.jp}

\begin{abstract}
We present six new and two previously published high-resolution
spectra of the quasar HS~1603+3820 ($\zem=2.542$) taken over an
interval of 4.2 years (1.2 years in the quasar rest frame). The
observations were made with the High-Dispersion Spectrograph on the
Subaru telescope and Medium-Resolution Spectrograph on the
Hobby-Eberly Telescope.  The purpose was to study the narrow
absorption lines (NALs). We use time variability as well as coverage
fraction analysis to separate intrinsic absorption lines, which are
physically related to the quasar, from intervening absorption
lines. By fitting models to the line profiles, we derive the
parameters of the respective absorbers as a function of time.  Only
the mini-BAL system at $\zabs\sim 2.43$ ($\voff\sim 9,500~\kms$) shows
both partial coverage and time variability, although two NAL systems
possibly show evidence of partial coverage.  We find that all the
troughs of the mini-BAL system vary in concert and its total
equivalent width variations resemble those of the coverage
fraction. However, no other correlations are seen between the
variations of different model parameters. Thus, the observed
variations cannot be reproduced by a simple change of ionization state
nor by motion of a homogeneous parcel of gas across the cylinder of
sight. We propose that the observed variations are a result of rapid
continuum fluctuations, coupled with coverage fraction fluctuations
caused by a clumpy screen of variable optical depth located between
the continuum source and the mini-BAL gas. An alternative explanation
is that the observed partial coverage signature is the result of
scattering of continuum photons around the absorber, thus the
equivalent width of the mini-BAL can vary as the intensity of the
scattered continuum changes.
\end{abstract}


\section{INTRODUCTION}
Quasars have been used as background sources to study the gaseous
phases of a variety of objects that are located along our sight-lines
to them.  These objects include not only {\it intervening} absorbers
such as intervening galaxies, the intergalactic medium (IGM), clouds
in the halo of the Milky Way, and the host galaxies of the quasars
themselves, but also {\it intrinsic} absorbers that are physically
associated with the quasar central engines. One of the most promising
candidates for the intrinsic absorbers are outflowing winds from the
quasars that could be accelerated by radiation pressure from the
accretion disk (Murray et al. 1995; Arav et al. 1995; Proga et
al. 2000) or by magnetocentrifugal forces (e.g., Everett 2005).
Outflowing winds are important components of quasar central engines
because they carry away angular momentum from the accretion disk and
allow the remaining gas in the disk to accrete onto the central black
hole. Quasar outflows are also important for cosmology since they
deliver energy, momentum, and metals to the interstellar and
intergalactic media, thus significantly affecting star formation and
galaxy assembly (e.g., Granato et al. 2004; Scannapieco \& Oh 2004;
Springel, Di Matteo \& Hernquist 2005).

Intrinsic narrow absorption lines (NALs; FWHM $\leq 500~\kms$) present
a powerful way to determine physical parameters of the accretion disk
winds.  Unlike the {\it broad} absorption lines (BALs; FWHM $\geq
2,000~\kms$; Weymann et al. 1991) that are also associated with
quasars outflows, NALs do not suffer from self-blending or from
saturation, so that line parameters are more easily evaluated. BALs
are thought to be associated with radiatively-driven outflows from
accretion disks (e.g., Weymann et al. 1991; Becker et al. 1997), while
it is difficult to separate intrinsic NALs from intervening NALs.
Mini-broad absorption lines (mini-BALs) are an intermediate subclass
between NALs and BALs, which are typically wider than those of NALs,
but narrower than BALs.  Mini-BALs have the advantages of both BALs
(i.e., high probability of being intrinsic lines) and NALs (i.e., line
profiles can be resolved into individual components), which make them
useful targets (e.g., Churchill et al. 1999; Hamann et al. 1997a;
Narayanan et al. 2004). BALs could probe the low-latitude, dense, fast
portion of the wind, while the mini-BALs and NALs may probe the
lower-density portion of the wind at high latitudes above the
disk. Thus, the study of mini-BALs and NALs complements the study of
BALs because the corresponding absorbers reside in different regions
of the outflow and allow us to sample different sets of physical
conditions.

To isolate intrinsic NALs, two tests are commonly used: (i) partial
coverage, i.e., trough dilution by un-occulted light (e.g., Hamann et
al.  1997b; Barlow \& Sargent 1997; Ganguly et al. 1999), and (ii)
variability of the absorption profiles within a few years in the
quasar rest frames (e.g., Hamann et al. 1997a; Narayanan et al. 2004;
Wise et al. 2004). These effects occur for intrinsic absorbers that
are very compact and very dense compared to intervening absorbers.
The fraction of quasars that host intrinsic \ion{C}{4} associated NALs
(NALs within 5,000~\kms\ of the quasar emission redshifts) has been
estimated to be $\sim 25$--27\%\ at $z < 2$, using the time
variability technique (Narayanan et al. 2004; Wise et al. 2004), and
$\sim$23\%\ at $z\sim 2.5$, based on partial coverage analysis (Misawa
et al. 2007). These signatures are a sufficient, but not a necessary,
condition to demonstrate the intrinsic nature of an absorber.  It is
very likely that some NALs without time variability or partial
coverage are also intrinsic.  

Both of the above methods have been applied to BAL and mini-BAL
systems. For example, partial coverage analysis has been carried out
by Petitjean \& Srianand (1999), Srianand \& Petitjean (2000,) Yuan et
al. (2002), and Ganguly et al. (2003), while time-variability has been
studied by Foltz et al. (1987), Turnshek et al. (1988), Smith \&
Penston (1988), Barlow et al. (1992), Barlow (1993), Vilkoviskij \&
Irwin (2001), Ma (2002), Narayanan et al. (2004), and Lundgren et
al. (2007). It is quite difficult (or almost impossible) to deblend
kinematic components in those heavily blending absorption features
using low/intermediate resolution spectra. However, some authors
successfully deblended narrower absorption components from each other
in such systems using high resolution spectra of $R > 40,000$ (e.g.,
Yuan et al. 2002; Ganguly et al. 2003). Systematic monitoring of
individual mini-BALs using high-resolution spectra taken on several
epochs can potentially provide very important constraints on the
properties of the absorbers.  However, no such campaign has been
attempted so far, to our knowledge.

The optically bright quasar HS~1603+3820 ($z_{em}$=2.542, B=15.9),
first discovered in the Hamburg/CfA Bright Quasar Survey (Hagen et
al. 1995; Dobrzycki et al. 1996), is known to have a large number (13)
of \ion{C}{4} doublets at $1.965 < \zabs < 2.554$ (Dobrzycki, Engels,
\& Hagen 1999). Using high-resolution spectra ($R=45,000$) taken with
the Subaru telescope, Misawa et al. (2003, 2005) classified all
\ion{C}{4} doublets into 8 \ion{C}{4} systems, and found that only a
mini-BAL system at $\zabs \sim 2.43$ (shift velocity\footnote{The
shift velocity is defined as positive for absorption lines that are
blueshifted from the quasar.}, $\voff = 8,300$--10,600~\kms) shows
both partial coverage and time variability in an interval of 0.36
years in the quasar rest frame. Based on these results, Misawa et
al. (2005) placed constraints (i) on the electron density ($n_e\geq
3.2\times 10^4~\cmmm$) and the absorber's distance from the quasar
($r\leq 6$~kpc), if a change in the ionization state causes the
variability, or (ii) on the time scale for the absorber to cross the
continuum source and the distance from the continuum source, if gas
motion across the background UV source causes the variability. In the
latter case, the crossing velocity is constrained to be $v_{cross}
\geq 8,000~\kms$, and the distance from the continuum source, $r \leq
0.2$~pc. This is larger than the size of the continuum source,
$R_{cont}\sim 0.02$~pc, but smaller than that of the BLR,
$R_{BLR}$$\sim$3~pc, estimated for this quasar.  On the other hand,
the other \ion{C}{4} NALs found towards the quasar show no sign of
being intrinsic.  Further monitoring observations make it possible to
(i) discriminate between changes in the ionization state and motion of
the absorbers across the continuum source as the cause of the time
variability seen in the \ion{C}{4} mini-BAL, and (ii) to confirm
whether the other \ion{C}{4} NALs really show neither partial coverage
nor variability.

In this paper, we present the results of monitoring HS~1603+3820 over
4.2 years (1.2 years in the quasar rest frame) based on eight spectra.
Six of these spectra were taken with the High Dispersion Spectrograph
(HDS; Noguchi et al. 2002) on the Subaru telescope, and two with the
Medium-Resolution Spectrograph (MRS; Horner, Engel, \& Ramsey, 1998)
on the Hobby-Eberly Telescope (HET). The first two spectra in this
time series were presented and discussed by Misawa et al (2003,
2005). With the new spectra we are able to sample the variations of
the absorption lines more densely and probe the internal structure of
the absorber.

In \S2, we describe the observations and data reduction. The methods
used for model fitting are outlined in \S3. The properties of the
mini-BAL and of the other NAL systems are examined in \S4. The
possible origins of the time variability seen in the mini-BAL system
are discussed in \S5, and a summary is given in \S6. We adopt $\zem =
2.542$ as the systemic redshift of the quasar, which was estimated
from its narrow emission lines (Misawa et al. 2003). Time intervals
between observations are given in the quasar rest frame throughout the
paper, unless otherwise noted.  We use a cosmology with $H_{0}$=75
\kms~Mpc$^{-1}$, $\Omega_{m}$=0.3, and $\Omega_{\Lambda}$=0.7.

\section{OBSERVATIONS AND DATA REDUCTION}

We observed HS~1603+3820 eight times over a period of 4.2 years in the
observed frame, from March 2002 to May 2006. We will call the dates of
these observations epoch~1 through epoch~8.  Six of the observations
were obtained with Subaru+HDS, using a 0{\farcs}8 or 1{\farcs}0 slit
width ($R = 45,000$ or 36,000), and adopting 2$\times$1 pixel binning
along the slit. The red grating, with a central wavelength of
4900~\AA, was used to cover as many \ion{C}{4} systems as possible,
except for the first and the last observations that used central
wavelengths of 6450~\AA\ and 5700~\AA. The other two spectra were
taken with the HET+MRS. The MRS features a pair of fibers: one for the
target and another for the sky, which enables us to perform sky
subtraction effectively. We used a 1{\farcs}5 fiber to get $R = 9,200$
spectra. A setup with a central wavelength of 7,000~\AA\ is required
for the queue observations with MRS+HET.

We reduced both the Subaru and HET data in a standard manner with the
IRAF software\footnote{IRAF is distributed by the National Optical
Astronomy Observatories, which are operated by the Association of
Universities for Research in Astronomy, Inc., under cooperative
agreement with the National Science Foundation.}. We used a Th-Ar
spectrum for wavelength calibration. We directly fitted the continuum,
which also includes substantial contributions from broad emission
lines\footnote{The broad emission lines probably do not vary between
observations, because bright quasars like HS~1603+3820 do not show
significant variability in magnitude on a time-scale as short as
$\sim$ 1 year (e.g, Giveon et al. 1999, Hawkins 2001, Kaspi et
al. 2007).}, with a third-order cubic spline function. Around heavily
absorbed regions, in which direct continuum fitting is difficult, we
used different techniques for the Subaru and HET spectra. For the
Subaru spectra, we adopted the interpolation technique introduced in
Misawa et al. (2003). Since the echelle blaze profile in HDS shows
time variability during a night, we cannot directly use the
instrumental blaze function as a continuum profile. In this scheme, we
obtain the continuum shape by interpolating between the two echelle
orders adjacent to the order of interest\footnote{If the order of
interest is $m$, we use orders $m-1$ and $m+1$ as reference
orders. However, in our data those adjacent orders are also somewhat
affected by broad absorption features. Therefore, we used orders $m-3$
and $m+3$ as reference orders.}, after weighting flux levels based on
the flat-frame spectrum [see eqn.~(1) of Misawa et al. 2003]. We then
divide the quasar spectrum by this interpolated continuum shape to
obtain the normalized spectrum. We have verified the validity of this
technique by applying it to a stellar spectrum; the continuum error is
always less than 3.3\%. For HET spectra, we used a blaze profile
constructed from the flat field spectrum as a continuum model after
multiplying it by a scaling factor to adjust its count level to that
of the quasar spectrum (the scaling factor is a ratio of counts in the
quasar spectrum to those in a flat spectrum for the same {\it
unabsorbed} region).

An observation log is given in Table~1, in which we list the
observation epoch, date of observation, relative time in the quasar
rest frame, instrument, wavelength coverage, total exposure time,
spectral resolving power, and signal-to-noise ratio (S/N) per pixel
(after rebinning to pixel scales of 0.03~\AA\ and 0.15~\AA\ for Subaru
and HET data, respectively). The S/N is evaluated around 5370~\AA,
close to the \ion{C}{4} mini-BAL. In Figure~1, we show a normalized
spectrum, after combining 6 spectra taken with Subaru+HDS in the
region redward of the \lya\ emission line of the quasar ($\lambda>
4300$~\AA).

\section{FITTING PROCEDURE}

We used the line-fitting software package {\sc minfit} (Churchill
1997; Churchill et al. 2003), with which we can fit absorption
profiles with four free parameters: redshift ($z$), column density
(\logn), Doppler parameter ($b$), and coverage fraction (\cf). Here,
the coverage fraction is the fraction of the background continuum
source and the broad-emission line region that is covered by the
intrinsic absorber.  The coverage fraction can be systematically
evaluated in an unbiased manner by considering the optical depth ratio
of resonant, rest-frame UV doublets of Lithium-like species (e.g.,
\ion{C}{4}, \ion{N}{5}, and \ion{Si}{4}), namely,
\begin{equation}
C_{f} = \frac{(R_{r}-1)^{2}}{1+R_{b}-2R_{r}}\; ,
\label{eqn:1}
\end{equation}
where $R_{b}$ and $R_{r}$ are the residual fluxes of the blue and red
members of the doublets in the continuum normalized spectrum (Hamann
et al. 1997b; Barlow \& Sargent 1997; Crenshaw et al. 1999).  A
$C_{f}$ value less than unity signifies that a portion of the
background source is not occulted by the absorber. This, in turn,
means the doublet is probably produced by an intrinsic absorber (e.g.,
Wampler et al. 1995; Barlow \& Sargent 1997).  We found 9 \ion{C}{4}
systems at $\zabs\ = 1.88$--2.55, as identified in Dobrzycki et
al. (1999), but we did not detect a \ion{C}{4} system at $\zabs =
2.5114$.  This system was probably a false detection (Dobrzycki 2005;
private communication). Among the 9 \ion{C}{4} systems, only the
system at $\zabs = 2.42$--2.45 is classified as a mini-BAL system,
because its line width is larger than 500~\kms.

We carried out model fits to all \ion{C}{4} systems in the same manner
as in our previous paper (Misawa et al. 2005), except for the
mini-BAL. Misawa et al. (2005) performed a manual Voigt profile fit to
the mini-BAL, because self-blending (blending of the blue and red
members of the doublet) occurs in this system. In its original form,
the {\sc minfit} code was not able to handle such self-blended
regions. The code was originally written for narrow absorption lines,
thus it fitted the models to the blue and red members of doublets {\it
separately}.  This procedure overproduced the residual intensity at
the self-blending regions because it added contributions from two
components. To avoid this problem, we revised the code so as to fit
models to the blue and red members of doublets {\it simultaneously},
i.e., by multiplying the contributions from the two doublet
members. In this paper, we fit the observed mini-BALs automatically
using this improved {\sc minfit}.

As in Misawa et al. (2005), we fitted only the first (bluest) two
\ion{C}{4} kinematic components in the system, at $\lambda_{obs} =
5284$--5318~\AA, because the other components are so heavily blended
with each other that we cannot separate them, even with the improved
{\sc minfit} (i.e., {\sc minfit} gives multiple solutions). At first,
we remove the contamination from \ion{Si}{2}~$\lambda$1527 in
System~B, at $\lambda = 5303$--5318~\AA. The \ion{Si}{2}~$\lambda$1527
line in the system did not show any detectable variability between the
first two observations. The \ion{C}{4} doublets for System~B show
neither partial coverage nor time variability, and are consistent with
it being an intervening system. Therefore, the
\ion{Si}{2}~$\lambda$1527 line is also not likely to vary. Thus, we
removed the contamination from \ion{Si}{2}~$\lambda$1527 by dividing
the observed spectra by a model profile synthesized with the line
parameters of \ion{Si}{2}~$\lambda$1527, determined from an average of
the first two observations presented in Misawa et al. (2005).

After removing the narrow \ion{Si}{2}~$\lambda$1527 components, we
applied the improved {\sc minfit} to the spectrum using two \ion{C}{4}
components, one narrow and one broad, as in Misawa et
al. (2005). During the fitting process, we initially forced the same
coverage fraction for the narrow and broad components, as in Misawa et
al. (2005).  In that paper we showed that, during the first two epochs
of observation, this solution produced an equally good fit as
solutions that allowed these coverage fractions to differ.  We also
considered the possibility of differing coverage fractions for the
narrow and broad components in later epochs of observation.

Figure~2 shows 6 spectra taken with Subaru+HDS, for which we find good
model fits. The fits themselves are shown in Figure~3.  We cannot fit
spectra obtained with HET+MRS in the same manner, because of their
lower resolution and S/N. Therefore, on the HET+MRS spectra, we
overplotted model profiles synthesized with parameters that are
linearly interpolated between the two adjacent epochs observed with
Subaru.  These models provide reasonably good fits to the HET+MRS
profiles.  The fitting parameters for the narrow and broad \ion{C}{4}
components, at different epochs, are summarized in Table~2. Columns
(1)--(3) give the epoch number, observation date and the relative time
in the quasar rest frame.  Columns (4)--(8) list, for the narrow
component, the shift velocity, \ion{C}{4} column density, \ion{C}{4}
Doppler parameter, coverage fraction, and observed-frame equivalent
width evaluated from the synthetic model spectrum. Columns (9)--(13)
are the same as columns (4)--(8), but for the broad component. We
confirmed that the fit parameters for the first two epochs are
consistent within 1$\sigma$ errors with those presented in Misawa et
al. (2005), based on a manual fitting procedure.

We also applied {\sc minfit} to all metal lines in the other NAL
systems. Since, as described later, we did not see any remarkable
variability in these NAL systems, we combined all 6 Subaru spectra to
produce a final spectrum with higher S/N ($\sim$96 per pixel) before
applying {\sc minfit} to the NALs\footnote{For system~G, we
combined only spectra in epochs~3, 7, and 8, because the other spectra
are affected by a data defect near the blue/red CCD gap.}. The best
fit parameters are summarized in Table~3. Column (1) is an
identification number for the line. Columns (2) and (3) give the
observed wavelength and redshift. Column (4) gives the shift velocity
from the quasar systemic redshift. Columns (5) and (6) give the column
density and the Doppler parameter with 1$\sigma$ errors. Column (7)
reports the coverage fraction with its 1$\sigma$ error (which includes
the error due to continuum level uncertainty as described
below). During a fitting trial, {\sc minfit} sometimes gives
unphysical coverage fractions such as $\cf < 0$ or $\cf \geq 1$. When
this happens, the other fit parameters such as the column density and
Doppler parameter have no physical meaning. These unphysical fit
results could be caused by errors in the continuum fit. Misawa et
al. (2005) found that the derived coverage fractions can be
significantly affected by continuum level errors, especially for very
weak components whose \cf\ values are close to 1. Therefore, if {\sc
minfit} gives unphysical \cf\ values for some components, we carry out
the fit again for the entire line, assuming $\cf = 1$ for those
components, following Misawa et al. (2007). In this case, we do not
evaluate the error in \cf. The best fit models are overlaid with the
observed spectra in Figure~4(a--i).

There are also other error sources for the fit parameters, such as
blending with other lines, and convolution of the true spectrum with
the instrumental line spread function (LSF). Both of these are
negligible if (1) the doublet is strong enough, i.e., $\log\; (N/\cmm)
\geq 14.0$ in the case of \ion{C}{4} doublets, (2) the normalized
residual flux is smaller than $\sim$0.5, and (3) the line profile is
much wider than the instrumental LSF (Misawa et al. 2005). Most
components in our sample are broad enough compared with the LSF
(corresponding to $b \sim 4~\kms$), and not severely blended with each
other, as described in section 4.2.

\section{VARIABILITY OF FIT PARAMETERS}

\subsection{The Mini-BAL System}

We previously reported that the \ion{C}{4} mini-BAL of System~A shows
trough dilution (Misawa et al. 2002), and variability within 0.36
years (Misawa et al. 2005). With the additional spectra of this paper,
we follow the variability of this system for 1.2 years in the quasar
rest frame, sampling it at eight epochs. \ion{N}{5} and \lya\ could
not be identified because of heavy contamination from the \lya\
forest. \ion{Si}{4} was not detected, even in our final, high-S/N
spectrum. Figure~5 presents graphically the evolution of the fit
parameters of the \ion{C}{4} mini-BAL, including (from top to bottom)
the total rest frame equivalent width at $\lambda = 5288$--5346~\AA,
the \ion{C}{4} column density, the Doppler parameter, the coverage
fraction, and the shift velocity. We discuss each in detail below.

The total rest frame equivalent width increased by about a factor of
two over a rest frame time interval of 0.36 years, and then decreased
over an interval of 0.6 years. During this entire period, the general
appearance of the line profile did not change significantly. Based on
the results of just the first two epochs, Misawa et al. (2005)
speculated that this mini-BAL may soon evolve into a BAL, with more
than 10\%\ of the flux absorbed over at least a 2,000~\kms\
range. This idea is not confirmed by the subsequent epochs, since the
mini-BAL was found to weaken. We also monitored the equivalent widths
of the narrow and broad components, separately. At first, both
components have almost the same equivalent width. But after epoch~3,
the equivalent width of the broad component was much higher than that
of narrow component.

The variability of the column density in the mini-BAL is complex.
This is especially true for the broad component, which showed a large
increase in column density within 0.13 years, from epoch~3 to epoch~7,
as shown in Figure~5. This increase in column density was concurrent
with the increase of equivalent width, as shown in the top window of
the figure. One possible explanation for this dramatic increase is a
change of the ionization state of the gas (Hamann et al. 1997b;
Narayanan et al. 2004), which we discuss further in \S5.

An interesting trend is seen in Doppler parameter variability. From
epoch~3 to epoch~5 (an interval of 0.09 years in the quasar rest
frame), the Doppler parameter of the broad component increased
suddenly from $\sim 350~\kms$ to $\sim 860~\kms$, and then it
decreased gradually. This abrupt jump is correlated with the sudden
increase of the column density in the broad component, as described
above. On the other hand, the narrow component did not show a
significant change in Doppler parameter.

The coverage fraction (assuming it is the same for the narrow and
broad components) shows a variability similar to that of the total
equivalent width.  After increasing to 0.45, the coverage fraction
decreased to 0.24. Inspection of the separate contributions of the
broad and narrow components to the total equivalent width shows that
the coverage fraction evolves similarly to the equivalent width of the
narrow component.  If the change of coverage fraction is caused by the
absorber's motion across the background flux source, the absorber size
is constrained to be smaller than the background source (about half of
the background source because of the maximum \cf\ value of $\sim$0.5).
At the same time, the column density increased, suggesting an
inhomogeneous absorber.  A more detailed discussion will be presented
in \S5.

The column density and equivalent width of the broad component
increased over the same period (epoch~3 to epoch~8) as its coverage
fraction decreased.  However, since we assumed that the broad and
narrow components had the same coverage fractions, we must assess
whether this assumption could lead to the observed trends.
Specifically, we consider the effect of assuming a constant \cf\ for
the broad component (adopting $\cf=0.36$ for epoch~3 through epoch~8,
and allowing the narrow component \cf\ to change as before).  This
model, which also produced an adequate fit to the data, is illustrated
with the dotted lines in the 2nd, 3rd, and 4th panels from the top of
Figure~5.  For this case, the column density of the broad component
does not increase as rapidly, but it still does increase.  The same
applies for the equivalent width.  Thus our conclusions regarding the
evolution of the column density and equivalent width are not
qualitatively changed by our assumption about the \cf\ value of the
broad component.

We have looked for correlated changes in all of the above parameters
of the mini-BAL profile by plotting them against each other. In
Figures~6 and 7, respectively, we show the relevant plots for the
narrow and broad components of the mini-BAL. Other than a tentative
trend between $C_f$ and $b$ in Figure~6 and between $C_f$ and $N_{\rm
C\; IV}$ in Figure~7, we see no convincing correlation. This lack of
correlated variability between model parameters is an important clue,
which we take into consideration in our later discussion.

The observed shift velocity represents the ejection velocity of the
absorber from the central engine projected along the line of
sight. While the line center of the narrow component remains nearly
constant, the broad component shifts by about 300~\kms\ ($\voff \sim
10,100~\kms \rightarrow 10,400~\kms$) from epoch~1 to epoch~7. If this
velocity shift is really due to the radiative pressure from the
continuum source, the acceleration of the gas can be calculated to be
$\sim$ 0.01~m~s$^{-2}$. However, the measurement of the center of the
broad component is highly uncertain, because the observed velocity
shift is much smaller than the total line width because of
self-blending of the blue and red members of the \ion{C}{4}
doublet. Moreover, the center of the broad component moved in both
directions (i.e., blueward and redward) during our
observations. Although it is still not clear if there is a truly
systematic increase in this parameter, we could regard the above value
as an upper limit on the acceleration of the absorber.

\subsection{NAL Systems}

In addition to the mini-BAL system, we detected 8 NAL systems in the
HS~1603+3820 spectra.  For these NALs, we do not detect significant
variability. Model fit results to the combined spectrum of the 6
Subaru observations (with a total exposure time of 8.9 hours and a S/N
per 0.03~\AA\ pixel of about 91 at $\lambda\sim 5370$~\AA) are
consistent with the results based on only the epoch~2 spectrum (a 1.7
hour exposure with $S/N \sim 49$ per pixel; Misawa et al. 2005), with
only a few small exceptions as described below. In this section we
discuss individual systems, other than system~A (the mini-BAL). The
profiles of the lines of these systems are shown on a common velocity
scale in Figure~4(a--i).

\begin{description}

\item[System B] --- We fit this strong \ion{C}{4} NAL, shown in
Figure~4a,b, with 15 components, although Misawa et al. (2005) used 19
components. The coverage fractions of two components, 11 and 14,
deviate from unity by more than 1$\sigma$, but both are heavily
blended with other components and their \cf\ values are not very
reliable. All components in the \ion{N}{5} NAL are consistent with
full coverage. Only the epoch~8 spectrum covered the \ion{Si}{4}
doublet of this system.  For the \ion{Si}{4} doublet, only component~7
suggests partial coverage, but the model fit is not good around the
left wing of this component. We conclude that this system does not
show strong evidence for partial coverage, and is probably an intervening
system, most likely a galaxy because of the strength of the
low-ionization absorption lines.

\item[System C] --- The \ion{C}{4} NAL in this system with a simple
absorption profile (see Figure~4c) is fitted with only one component,
as in Misawa et al. (2005). Both fits are consistent with full
coverage.  We also did not find any evidence for time variability.
Except for its small shift velocity, $\sim 430~\kms$, there is no
evidence to suggest that this system is intrinsic.

\item[System D] --- Misawa et al. (2005) fitted this system's
\ion{C}{4} and \ion{Si}{4} profiles with 4 and 2 components,
respectively, and did not find any partial coverage. However, we found
an additional \ion{C}{4} component at $\voff = -160~\kms$ from the
system center that shows partial coverage, and we have no reason to
suspect systematic error.  The line profiles are shown in Figure~4d.
This component was not studied by Misawa et al. (2005) because of a
data defect. We also now find that component~1 in the \ion{Si}{4} NAL
possibly shows partial coverage, although it was consistent with full
coverage in the shorter exposure presented by Misawa et al. (2005).
We also note that it is surprising that the \ion{C}{4}~$\lambda$1548
is black at the position of this \ion{Si}{4} component if the partial
coverage is true. However, the fact that partial coverage is apparent
for both a \ion{Si}{4} and a \ion{C}{4} component suggests that this
system may be an intrinsic system. The fact that this system is
redshifted from the quasar systemic redshift also supports this
idea. Although this system is located near the top of the \ion{C}{4}
emission line, there is no residual flux at the position of the
\ion{C}{4} doublet for components~4 and 5.  These components must
therefore cover both the continuum source and broad emission line
region.  We do not see any variability over the course of our
observations.

\item[System E] --- Misawa et al. (2005) fitted the \ion{C}{4} NAL
with only two components, and one of them showed evidence of partial
coverage. However, our higher-S/N spectrum (shown in Figure~4e)
requires 5 components to fit the \ion{C}{4} NAL, and three of them are
consistent with full coverage. Although components~3 and 5 deviate
from full coverage, they are heavily blended with the other
components, so this result is quite uncertain.  The shift velocity of
the system is also extremely large, $\sim$60,000~\kms. There is no
compelling reason to believe that this system is intrinsic to the
quasar.

\item[System F] --- Our best-fitting model is almost consistent with the
previous one, having the same number of components at almost the same
positions (see Figure~4f). All components are consistent with full
coverage except for component~1 that is blended with the strong
component~2. This system also has a very large shift velocity, $\sim
50,000~\kms$. We consider it probable that this is an intervening
system.

\item[System G] --- Because the \ion{C}{4} NAL in this system was
positioned at the edge of the blue CCD of the HDS (around which bad
pixels are common) in some of our observing setups, the absorption
profiles are slightly or severely different in some
spectra. Therefore, we combined only spectra from epochs~3, 7, and 8
(these are not affected by the data defect) to produce a final
spectrum for this system (Figure~4g). We fit the \ion{C}{4} NAL with 6
components, and all of them are consistent with full
coverage. Component~5 was not fitted by Misawa et al. (2005) because
of the unphysical line ratio of the blue and red component, due to
line blending. This system is probably an intervening system.

\item[System H] --- Two components are necessary to fit this
\ion{C}{4} NAL (see Figure~4h).  The shallower absorption feature,
centered on component~2, has an unphysical ratio of its blue and red
members. Probably the blue member is contaminated by other lines. Only
component~1 is useful, and it yields $\cf=1$. In this system, we again
do not see any evidence of intrinsic properties.

\item[System I] --- This system was detected only in the epoch~8
spectrum, because it was not covered in the other epochs. We used 7
components to fit the \ion{C}{4} NAL (see Figure~4i).  Component~4 may
show partial coverage, but its \cf\ value differs from unity by only
slightly more than 1$\sigma$.  The case for partial coverage for
component~6 is somewhat more compelling, since the {\sc minfit}
fitting procedure yielded a value $\cf\ = 0.87 \pm 0.02$.  Component~4
is not blended with other components, and component~6 is only weakly
blended.  Although the S/N for this spectrum is not very high in the
relevant region, this system may be another candidate for an intrinsic
system.

\end{description}

\section{DISCUSSION: THE ORIGIN OF TIME VARIABILITY}

In view of the observational constraints derived above, we evaluate
here different scenarios for the origin of the variability of the
\ion{C}{4} mini-BAL.  A fundamental assumption underlying our
discussion is that the kinematic components making up system~A
represent parcels of absorbing gas at different positions along the
line of sight. In effect, we assume that the absorber is embedded in
an accelerating outflow from the quasar, thus the blueshift of a
kinematic component increases with its distance from the quasar
continuum source.

A very important observational clue from the data presented above is
the fact that all the kinematic components of the mini-BAL system vary
in concert (i.e., the depths of all troughs increase or decrease
together). These changes appear to be driven largely by changes in the
coverage fraction, as shown in Figure~5. Other physical parameters
($N_{\rm C\; IV}$ and $b$) are varying as well, but there does not
seem to be an obvious correlation between changes of different profile
parameters, at least in the bluest troughs of the mini-BAL (see
Figures 6 and 7).

\subsection{Transverse Motion of a Homogeneous Absorber}

In Misawa et al. (2005) we favored an interpretation in which motion
of the (homogeneous) absorber across the line (or cylinder) of sight
causes the observed variability (see also Hamann et al. 1997a). The
assumed scenario is depicted in the cartoon of Figure~2a of Hamann \&
Sabra (2004).  With time, an increasing fraction of the cylinder of
sight was covered by the absorber, leading to changes in the coverage
fraction. In this context and based on two snapshots of the spectrum,
we derived a transverse velocity of 8,000~\kms\ and a constraint of $r
< 0.2$~pc on the distance of the absorber from the continuum source
(see details in Misawa et al. 2005).

That model was motivated largely by the observation that the coverage
fraction was the only quantity that changed substantially between the
first two observations. The subsequent observations show that scenario
to be too simple, namely the column density of the absorber changes
with time as well, suggesting that the absorber is inhomogeneous and
its internal structure plays an important role in the variability of
the mini-BAL profile. Thus, a more plausible scenario is the one
depicted in Figure~2b of Hamann \& Sabra (2004) and in Figure 5 of
de~Kool, Korista, \& Arav (2002), where the absorbing medium is
inhomogeneous, with clumps spanning a range of column densities
scattered over the cylinder of sight.

There is a remaining difficulty with this scenario, however.  The fact
that all the troughs in the mini-BAL vary in concert would require a
coincidence between the motions of the individual clumps. More
specifically, the projected areas of all the clumps within the
cylinder of sight should vary in the same way (i.e., rise and fall
with time in the same manner for all clumps). We consider such a
coordinated sequence of variations rather unlikely, therefore, we
regard this interpretation as implausible. We note that if future
observations show this pattern to be common, then this interpretation
can be conclusively rejected.

\subsection{Internal Motions in the Absorbing Gas}

At first glance, compression of the absorber might be able to produce
a trend similar to that of Figure~7b, where the coverage fraction
decreases as the column density increases in one of the kinematic
components of the mini-BAL. However, on closer examination, such a
compression would require highly supersonic motions within the
absorber. Assuming the background continuum source size is 0.02~pc
(this is the UV-emitting region of the accretion disk; see Misawa et
al. 2005), a change in the coverage fraction from 0.36 to 0.27 within
0.13 years (epochs 5 and 7) implies a speed of order $10^4~\kms$. In
comparison, the speed of sound in a $\sim10^4$~K gas is only
$10~\kms$. Thus we consider such a scenario untenable.

\subsection{Scattering of Continuum Photons Around the Absorber}

The observed partial coverage signature can, in principle, be
explained if continuum photons are re-directed by a scattering medium
towards the observer, when they were initially traveling in a
different direction. Thus, the absorber can be homogeneous and its
projected size need not be smaller than the cylinder of sight towards
the continuum source. The dilution of the troughs by photons scattered
around the absorber can change with time as the absorber moves
relative to the scattering medium or as the conditions in the
scattering medium change. In the context of this scenario, the {\it
apparent} coverage fraction can vary according to the behavior of the
scattering medium. Moreover, the variations of the coverage fraction
need not be tied in any way to variations of other parameters of the
absorber, such as the \ion{C}{4} column density, the $b$-parameter, or
the shift velocity. This idea can be tested observationally through
spectropolarimetry. In particular, we would expect the fractional
polarization in the absorption mini-BAL troughs to be higher than in
the continuum (see the analogous test in BAL quasars by Brotherton et
al. 1997, Ogle et al. 1999, and Lamy \& Hutsem\'ekers 2000, for
example).  Furthermore, we would also expect this fractional
polarization to increase as the coverage fraction decreases.

\subsection{Change of Ionization State of the Absorbing Gas}

Changes in the level of the ionizing continuum can also bring about
absorption-line variability since they change the relative ionic
populations in the absorber (in this particular case, the relative
populations of \ion{C}{3}, \ion{C}{4} and \ion{C}{5}). In principle, a
monotonic change in the continuum intensity can produce a
non-monotonic change in the absorption line strength. For example, an
absorber initially at a low ionization state that sees a rising
ionizing flux will first respond with an increase in the strength of
its \ion{C}{4} absorption lines (as
\ion{C}{3}$\;\rightarrow\;$\ion{C}{4}) and then with a decrease (as
\ion{C}{4}$\;\rightarrow\;$\ion{C}{5}). The variability time scale of
the absorption lines provides an upper limit on the ionization or
recombination time of the absorbing gas, hence its electron density
(see the detailed discussion in Hamann et al. 1997b and Narayanan et
al. 2004).  The density limit can be obtained from $n_e \geq (\alpha_r
t_{var})^{-1}$, where $\alpha_r$ is a recombination coefficient and
$t_{var}$ is a variability time scale. This is a simplified picture in
which the gas is in ionization equilibrium and \ion{C}{4} is the
dominant ionic species of C.

Since the \ion{C}{4} ionic column densities have fluctuated up and
down {\it more than once} over the course of our monitoring
observations, the ionizing continuum seen by the absorber must be
fluctuating too. Moreover, the continuum intensity must be fluctuating
by at least a factor of 3 on rest-frame time scales of 6 months or
less to produce the observed changes in the \ion{C}{4} ionic column
density (see the calculations of Hamann 1997).  Taking this
interpretation at face value (but see discussion below), we can use
the shortest variability time scale that we have observed (16 days in
the quasar rest frame, epochs 5$\;\rightarrow\;$7; see Fig.~5) to
constrain the density of the gas. We consider both an increase and a
decrease in the ionization state of the gas as a cause for the
observed change in the \ion{C}{4} column density (i.e.,
\ion{C}{3}$\;\rightarrow\;$\ion{C}{4} or
\ion{C}{5}$\;\rightarrow\;$\ion{C}{4}), with respective recombination
coefficients of $5.3\times 10^{-12}~{\rm cm^3~s^{-1}}$ and $2.8\times
10 ^{-12}~{\rm cm^3~s^{-1}}$ (from Arnaud \& Rothenflugh 1986;
assuming a nominal gas temperature of 20,000~K, following Hamann
1997). Thus we obtain a limit on the electron density of $n_e >
1\times 10^5~\cmmm$, using the methodology of Narayanan et al. (2004)
and a bolometric luminosity of $L_{bol} = 2.5\times 10^{48}~{\rm
erg~s}^{-1}$ (Misawa et al. 2005). To convert this limit to a limit on
the distance of the absorber from the continuum source, $r$, we must
assume a value of the ionization parameter.  If the \ion{C}{4} column
density fluctuations reflect conversions of
\ion{C}{3}$\;\leftrightarrow\;$\ion{C}{4} in a lower-ionization
absorber, then the calculations of Hamann (1997) indicate an
ionization parameter of $U\sim 0.002$ and lead to $r<8$~kpc. If, on
the other hand, the \ion{C}{4} column density fluctuations reflect
conversions of \ion{C}{5}$\;\leftrightarrow\;$\ion{C}{4} in a
higher-ionization absorber, then the calculations of Hamann (1997)
indicate an ionization parameter of $U\sim 0.06$ and lead to
$r<1$~kpc. In this context the simultaneous variability of all the
mini-BAL troughs implies that these troughs should represent gas
parcels of similar densities for the following reason. Since all
absorbing gas parcels must be within the cylinder of sight to the
continuum source, the time lag between changes in the continuum
intensity at the source and our observation of a change in the
mini-BAL profile is the sum of the light travel times from the
continuum source to the absorber and from the absorber to the
observer, plus the recombination time of the gas. The sum of light
travel times is the same regardless of the location of a gas parcel
{\it along} the cylinder of sight, therefore, a difference in time lag
can only result from different recombination times, i.e., different
gas densities. We can then conclude that the density does not vary
significantly across the line profile, which implies in turn that the
shape of the mini-BAL profile is determined primarily by the
combination of column density (i.e., physical thickness) and coverage
fraction as a function of velocity.

Fluctuations in the ionization state of the absorber are an attractive
explanation of the observed variations in the strength of the mini-BAL
in HS~1603+3820. However, the mini-BAL appears to vary much faster than
any expected variations in the UV continuum of such a luminous quasar
(see for example, Giveon et al. 1999, Hawkins 2001, and Kaspi et
al. 2007). Moreover, Fleming \& Kennefick (2006) found no variability
of this particular quasar in 10 days in the observed frame.
\footnote{Unfortunately, our spectra cannot provide information on
variations of the continuum level because we cannot calibrate their
flux scale. This is a consequence of the fact that the blaze function
of the HDS on Subaru changes with time over the course of a night.}
Therefore we propose that the variations of the ionizing continuum
{\it seen by the absorber} are caused by a screen of variable optical
depth between it and the (relatively steady) continuum source. This
screen could be the ionized, clumpy, inner part of the outflow, i.e.,
the shielding or hitchhiking gas invoked in the outflow models of
Murray et al. (1995), and arising naturally in the more detailed
calculations of Proga et al. (2000). The two-dimensional, axisymmetric
models of Proga et al. (2000) also show that disk winds can be
accelerated up to 15,000~\kms, and that they are unsteady and generate
dense knots periodically. The time variability seen in the mini-BAL
system of HS~1603+3820 may be related to this mechanism. The
conditions in the screen may be similar to those of ``warm''
absorbers, observed in the X-ray spectra of many Seyfert galaxies,
where it often co-exists with a UV absorber (e.g., Crenshaw et
al. 1999).  We note, by way of example, that models for the warm
absorbers in NGC~5548, IRAS~1339+2438, and NGC~3783 require multiple
ionization phases for an adequate fit to the X-ray spectrum (Kaastra
et al. 2000, Sako et al. 2001, Blustin et al. 2002), one of which may
fulfill the requirements for the inner screen.

If the optical depth of the screen at wavelengths around the
\ion{C}{4} edge is close to unity, then small fluctuations in the
column density can cause large changes in the shape and intensity of
the transmitted ionizing continuum (e.g., Fig.~6 of Murray et
al. 1995, or Fig.~1 of Hamann 1995), leading to the observed changes
in the mini-BAL of HS~1603+3820.  This may be analogous to the
best-studied warm absorber, in the Seyfert galaxy NGC~3783, where
fluctuations in the ionization structure of the absorbing gas have
been observed over the course of 6 months (e.g., Netzer et
al. 2003). There is also evidence that the opacity of the warm
absorber in NGC~3783 fluctuates over time intervals as short as a
month, in response to variations in the X-ray continuum by a factor of
2 (Krongold et al. 2005).  Within this context we can also understand
why the variations in the coverage fraction of the mini-BAL gas are
unrelated to variations of its column density. Because both media can
be clumpy and since differential rotation of the wind causes the inner
screen to move faster than the mini-BAL gas across the cylinder of
sight, the intensity of the continuum transmitted through the screen
and received by the mini-BAL gas would vary with time.  Moreover, the
coverage fraction need not be correlated with the intensity of the
transmitted continuum.

We have considered the possibility that one of the observed NAL
systems at smaller blueshifts relative to the quasar could represent
the inner screen. However, none of the three candidate \ion{C}{4}
systems, B, C, and D, can satisfy the requirements for the screen.
The very fact that they have \ion{C}{4} absorption lines suggests that
their ionization state is too low, and moreover, systems B and D have
even lower-ionization lines. In addition, none of the three systems
are variable, which is necessary in order for the screen to perform
the desired function. More promising ways of finding the screen
signature are X-ray observations in search of the classic warm
absorber feature, \ion{O}{7} and \ion{O}{8} edges. Thus, we will be
able to determine which of the \lya\ absorption lines seen at shorter
wavelengths might be associated with the screen material.

\section{SUMMARY AND CONCLUSIONS}

We have been monitoring the absorption lines in the spectrum of the
quasar HS~1603+3820 since 2002, using the Subaru and Hobby-Eberly
Telescopes. We have obtained six Subaru+HDS and two HET+MRS spectra
spanning an interval of approximately 1.2 years in the quasar rest
frame and probing rest-frame time scales as short as 16 days.  We have
determined the physical parameters of the 9 absorption-line systems by
fitting Voigt models to the line profiles. Our main results are as
follows:

\begin{enumerate}

\item
We detect one \ion{C}{4} mini-BAL and eight \ion{C}{4} NAL systems in
the quasar spectrum.  The partial coverage signature and time
variability of the mini-BAL show it to be intrinsic to the quasar.
Two of the NAL systems (D and I) may also be intrinsic, based upon our
partial coverage analysis.

\item
We are able to fit models only to the bluest portion of the \ion{C}{4}
mini-BAL profile where self-blending is not severe.  All fit
parameters (i.e., column density, Doppler parameter, coverage
fraction, and shift velocity) as well as the total equivalent width of
the system vary significantly with time, even on short time
scales. However, the profile parameters do not appear to change in
concert with each other, with one exception: the equivalent widths of
all the troughs in this system vary together and approximately follow
the variations of the coverage fraction determined from the model
fits.

\item
We have examined a number of ways of explaining the above variations
of the \ion{C}{4} mini-BAL and we have found two viable possibilities.
The first possibility is that the observed partial coverage signature
is the result of continuum photons scattering around the absorber and
into our cylinder of sight. The observed changes in mini-BAL
equivalent widths are thus produced by variations in the scattered
continuum that dilutes the absorption troughs.  In the second
possibility, the illumination of the UV absorber fluctuates on short
time scales by a factor of up to 3. We suggest that these fluctuations
are caused by a screen of variable optical depth between the mini-BAL
gas and the continuum source. This screen might be identified with the
shielding gas invoked or predicted in some outflow models.  Moreover,
it could be analogous to the ``warm'' absorbers observed in the X-ray
spectra of Seyfert galaxies and some quasars. This picture can also
explain the variations in the coverage fraction, which appear to be
unrelated to the ionic column density changes.

\end{enumerate}

\acknowledgments This work was supported by NASA grant NAG5-10817. We
are grateful to the staff of the Subaru telescope, which is operated
by the National Astronomical Observatory of Japan. We would also like
to thank Christopher Churchill for providing us with the {\sc minfit}
software package. ME acknowledges partial support from the Theoretical
Astrophysics Visitors' Fund at Northwestern University and thanks the
members of the theoretical astrophysics group for their warm
hospitality.  The Hobby-Eberly Telescope (HET) is a joint project of
the University of Texas at Austin, the Pennsylvania State University,
Stanford University, Ludwig-Maximillians-Universit\"at M\"unchen, and
Georg-August-Universit\"at G\"ottingen. The HET is named in honor of
its principal benefactors, William P. Hobby and Robert E. Eberly.  We
also thank the anonymous referee for helpful comments and suggestions.

\clearpage

\begin{deluxetable}{clcllcrc}
\tabletypesize{\scriptsize}
\setcounter{table}{0}
\tablecaption{Log of Monitoring Observations \label{t1}}
\tablewidth{0pt}
\tablehead{
\colhead{}                       &
\colhead{}                       & 
\colhead{Time\tablenotemark{a}}  & 
\colhead{}                       & 
\colhead{Wavelength Coverage}    & 
\colhead{Exposure}               &
\colhead{Resolving}              &
\colhead{}                       \\
\colhead{Epoch}                  &
\colhead{Date}                   & 
\colhead{(years)}                &
\colhead{Instrument}             & 
\colhead{(\AA)}                  & 
\colhead{(s)}                    & 
\colhead{Power}                  & 
\colhead{$S/N$\tablenotemark{b}} \\
\colhead{(1)} &
\colhead{(2)} &
\colhead{(3)} &
\colhead{(4)} &
\colhead{(5)} &
\colhead{(6)} &
\colhead{(7)} &
\colhead{(8)} 
}
\startdata
1\tablenotemark{c} & 2002 Mar 23          & 0.00 & Subaru+HDS & 5080--6420             & 2700 & 45,000 & 46.5 \\
2\tablenotemark{c} & 2003 Jul  7          & 0.36 & Subaru+HDS & 3520--4850, 4930--6260 & 6000 & 45,000 & 48.7 \\
3                  & 2005 Feb 26          & 0.83 & Subaru+HDS & 3520--4855, 4925--6275 & 7100 & 36,000 & 43.0 \\
4                  & 2005 May 10          & 0.89 &    HET+MRS & 4500--6200             & 1800 &  9,200 & 17.4 \\
5                  & 2005 Jun 29          & 0.92 & Subaru+HDS & 3530--4855, 4925--6280 & 3600 & 45,000 & 31.1 \\
6                  & 2005 Aug  3,8        & 0.95 &    HET+MRS & 4500--6170             & 4000 &  9,200 & 28.5 \\
7                  & 2005 Aug 19          & 0.96 & Subaru+HDS & 3520--4850, 4920--6235 & 3600 & 36,000 & 43.6 \\
8                  & 2006 May 31--Jun 1   & 1.18 & Subaru+HDS & 4320--5630, 5740--7050 & 9000 & 45,000 & 52.5 \\
\enddata
\tablenotetext{a}{The relative time since the first observation, in the quasar rest frame.}
\tablenotetext{b}{S/N per bin around 5370~\AA\ after rebinning.
                  The final bin size is 0.03~\AA\ for the Subaru+HDS
                  spectra and 0.15~\AA\ for the HET+MRS spectra.}
\tablenotetext{c}{These spectra we originally presented and discussed
                  by Misawa et al. (2003, 2005).}
\end{deluxetable}
\clearpage

\begin{deluxetable}{clccccccccccccc}
\rotate
\tabletypesize{\scriptsize}
\tablecaption{Best fit parameters of C IV mini-BAL \label{t2}}
\tablewidth{0pt}
\tablehead{
\colhead{} &
\colhead{} &
\colhead{} &
\multicolumn{5}{c}{narrow component} &
\colhead{} &
\multicolumn{5}{c}{broad component} \\
\cline{4-8} 
\cline{10-14} \\
\noalign{\vskip -6pt}
\colhead{} & 
\colhead{} & 
\colhead{Time\tablenotemark{a}} & 
\colhead{\voff} & 
\colhead{log$N$} &
\colhead{$b$} &
\colhead{\cf} &
\colhead{$EW^b$} &
\colhead{} &
\colhead{\voff} & 
\colhead{log$N$} &
\colhead{$b$} &
\colhead{\cf} &
\colhead{$EW^b$} \\
\colhead{Epoch} & 
\colhead{Date} & 
\colhead{(yrs)} & 
\colhead{(\kms)} & 
\colhead{(\cmm)} & 
\colhead{(\kms)} & 
\colhead{} &
\colhead{(\AA)} &
\colhead{} &
\colhead{(\kms)} & 
\colhead{(\cmm)} & 
\colhead{(\kms)} & 
\colhead{} &
\colhead{(\AA)} \\
\colhead{(1)}  &
\colhead{(2)}  &
\colhead{(3)}  &
\colhead{(4)}  &
\colhead{(5)}  &
\colhead{(6)}  &
\colhead{(7)}  &
\colhead{(8)}  &
\colhead{}     &
\colhead{(9)}  &
\colhead{(10)} &
\colhead{(11)} &
\colhead{(12)} &
\colhead{(13)} 
}
\startdata
1 & 2002 Mar 23        & 0.00 & 10648 & 14.55$\pm$0.07 & 85$\pm$3 & 0.32$\pm$0.03 & 1.49 & & 10118 & 14.53$\pm$0.04 & 233$\pm$ 6 & 0.32$\pm$0.03 & 1.88 \\
2 & 2003 Jul  7        & 0.36 & 10637 & 14.72$\pm$0.07 & 98$\pm$5 & 0.45$\pm$0.04 & 2.79 & & 10273 & 14.64$\pm$0.05 & 274$\pm$15 & 0.45$\pm$0.04 & 3.32 \\
3 & 2005 Feb 26        & 0.83 & 10644 & 14.75$\pm$0.03 & 84$\pm$2 & 0.36$\pm$0.01 & 2.15 & & 10309 & 14.53$\pm$0.03 & 348$\pm$15 & 0.36$\pm$0.01 & 2.22 \\
4 & 2005 May 10 $^c$   & 0.89 & 10643 & 14.70          & 81       & 0.33          & 1.83 & & 10259 & 14.67          & 692        & 0.33          & 2.86 \\
5 & 2005 Jun 29        & 0.92 & 10642 & 14.68$\pm$0.05 & 80$\pm$3 & 0.32$\pm$0.02 & 1.72 & & 10235 & 14.74$\pm$0.06 & 863$\pm$99 & 0.32$\pm$0.02 & 3.20 \\
6 & 2005 Aug 3--8 $^d$ & 0.95 & 10641 & 14.72          & 71       & 0.28          & 1.49 & & 10345 & 14.82          & 793        & 0.28          & 3.29 \\
7 & 2005 Aug 19        & 0.96 & 10641 & 14.74$\pm$0.03 & 69$\pm$2 & 0.27$\pm$0.01 & 1.43 & & 10381 & 14.85$\pm$0.02 & 770$\pm$31 & 0.27$\pm$0.01 & 3.36 \\
8 & 2006 May 31--Jun 1 & 1.18 & 10637 & 14.76$\pm$0.03 & 71$\pm$2 & 0.24$\pm$0.01 & 1.32 & & 10291 & 14.90$\pm$0.02 & 587$\pm$18 & 0.24$\pm$0.01 & 3.27 \\
\enddata
\tablenotetext{a}{The relative time since the first observation, 
                  in the quasar rest frame.}
\tablenotetext{b}{Equivalent width of broad/narrow component in the
                  rest frame that is evaluated with the synthesized
                  spectrum with fit parameters.}
\tablenotetext{c}{Interpolated value, using the data on 2005 February 26
                  and 2005 June 29.}
\tablenotetext{d}{Interpolated value, using the data on 2005 June 29
                  and 2005 August 19.}
\end{deluxetable}
\clearpage

\begin{deluxetable}{lcccccc}
\tabletypesize{\scriptsize}
\tablecaption{Metal lines in the combined spectrum \label{t3}}
\tablewidth{0pt}
\tablehead{
\colhead{Line ID}         & 
\colhead{$\lambda_{obs}$} & 
\colhead{$z_{abs}$}       & 
\colhead{\voff}           &
\colhead{$\log\; N$}      &
\colhead{$b$}             &
\colhead{$C_{f}$}         \\
\colhead{}                & 
\colhead{(\AA)}           & 
\colhead{}                & 
\colhead{(km s$^{-1}$)}   & 
\colhead{(cm$^{-2}$)}     &
\colhead{(km s$^{-1}$)}   & 
\colhead{}                \\
\colhead{(1)} &
\colhead{(2)} &
\colhead{(3)} &
\colhead{(4)} &
\colhead{(5)} &
\colhead{(6)} &
\colhead{(7)} 
}
\startdata
\tableline		      			      
\multicolumn{7}{c}{System B : $z_{abs}=2.4785$}       
\\			      			      
\tableline
\multicolumn{7}{c}{\ion{C}{4} $\lambda$1548 / \ion{C}{4} $\lambda$1551 ($W_{obs}$ = 4.58 / 3.32 \AA)}                    \\
 1............        & 5381.8 & 2.47618 & 5627 & 13.36 $\pm$ 0.01 &  8.2 $\pm$ 0.2 & 1.00                   \\
 2............        & 5382.2 & 2.47642 & 5606 & 14.16 $\pm$ 0.03 &  6.3 $\pm$ 0.2 & 1.00                   \\
 3............        & 5382.8 & 2.47685 & 5569 & 13.52 $\pm$ 0.01 & 22.4 $\pm$ 0.4 & 1.00                   \\
 4............        & 5383.8 & 2.47749 & 5514 & 13.83 $\pm$ 0.00 & 15.6 $\pm$ 0.1 & 1.00                   \\
 5............        & 5383.9 & 2.47752 & 5511 & 14.69 $\pm$ 0.46 &  5.5 $\pm$ 1.1 & 1.00$^{+0.03}_{-0.03}$ \\
 6............        & 5384.3 & 2.47779 & 5488 & 12.87 $\pm$ 0.29 &  4.1 $\pm$ 2.1 & 0.90$^{+0.60}_{-0.40}$ \\
 7............        & 5384.5 & 2.47793 & 5476 & 13.58 $\pm$ 0.01 &  6.7 $\pm$ 0.2 & 1.00                   \\
 8............        & 5385.0 & 2.47826 & 5447 & 14.06 $\pm$ 0.00 & 14.8 $\pm$ 0.2 & 1.00                   \\
 9............        & 5385.6 & 2.47862 & 5416 & 13.96 $\pm$ 0.10 & 14.7 $\pm$ 2.2 & 0.97$^{+0.03}_{-0.03}$ \\
 10...........        & 5386.2 & 2.47899 & 5384 & 14.13 $\pm$ 0.00 & 27.1 $\pm$ 0.2 & 1.00                   \\
 11...........        & 5386.2 & 2.47899 & 5384 & 13.62 $\pm$ 0.21 &  5.6 $\pm$ 1.2 & 0.76$^{+0.18}_{-0.18}$ \\
 12...........        & 5387.1 & 2.47959 & 5333 & 12.96 $\pm$ 0.21 &  7.3 $\pm$ 1.0 & 1.00$^{+0.66}_{-0.43}$ \\
 13...........        & 5387.4 & 2.47977 & 5317 & 13.74 $\pm$ 0.01 &  7.8 $\pm$ 0.1 & 1.00                   \\
 14...........        & 5388.2 & 2.48028 & 5273 & 13.56 $\pm$ 0.18 & 28.1 $\pm$ 2.2 & 0.45$^{+0.49}_{-0.17}$ \\
 15...........        & 5388.6 & 2.48055 & 5250 & 14.03 $\pm$ 0.03 &  4.7 $\pm$ 0.1 & 1.00                   \\
\multicolumn{7}{c}{\ion{N}{5} $\lambda$1239 / \ion{N}{5} $\lambda$1243   ($W_{obs}$ = 0.35 / 0.19 \AA)}                    \\
 1............        & 4308.0 & 2.47748 & 5514 & 13.26 $\pm$ 0.01 &  7.7 $\pm$ 0.3 & 1.00                   \\
 2............        & 4308.4 & 2.47784 & 5483 & 12.73 $\pm$ 0.06 &  9.4 $\pm$ 1.5 & 1.00                   \\
 3............        & 4308.9 & 2.47824 & 5449 & 13.13 $\pm$ 0.06 & 19.7 $\pm$ 2.7 & 1.00                   \\
 4............        & 4309.6 & 2.47882 & 5399 & 13.22 $\pm$ 0.05 & 28.2 $\pm$ 4.0 & 1.00                   \\
\multicolumn{7}{c}{\ion{Si}{4} $\lambda$1394 / \ion{Si}{4} $\lambda$1403 ($W_{obs}$ = 1.59 / 1.05 \AA)}                  \\
 1............        & 4845.2 & 2.47635 & 5612 & 12.44 $\pm$ 0.03 & 15.5 $\pm$ 0.8 & 1.00                   \\
 2............        & 4845.3 & 2.47642 & 5606 & 13.20 $\pm$ 0.01 &  4.0 $\pm$ 0.1 & 1.00                   \\
 3............        & 4846.8 & 2.47751 & 5512 & 12.49 $\pm$ 0.01 &  7.5 $\pm$ 0.2 & 1.00                   \\
 4............        & 4847.4 & 2.47793 & 5476 & 12.72 $\pm$ 0.01 &  2.9 $\pm$ 0.1 & 1.00                   \\
 5............        & 4847.8 & 2.47824 & 5449 & 13.05 $\pm$ 0.10 & 21.3 $\pm$ 0.9 & 0.94$^{+0.29}_{-0.21}$ \\
 6............        & 4848.3 & 2.47860 & 5418 & 13.12 $\pm$ 0.00 &  9.3 $\pm$ 0.1 & 1.00                   \\
 7............        & 4848.9 & 2.47899 & 5384 & 13.05 $\pm$ 0.07 & 11.9 $\pm$ 0.3 & 0.69$^{+0.08}_{-0.07}$ \\
 8............        & 4849.7 & 2.47960 & 5332 & 12.98 $\pm$ 0.01 & 27.7 $\pm$ 0.4 & 1.00                   \\
 9............        & 4849.9 & 2.47976 & 5318 & 12.86 $\pm$ 0.01 &  4.6 $\pm$ 0.1 & 1.00                   \\
 10...........        & 4851.0 & 2.48054 & 5251 & 13.39 $\pm$ 0.01 &  4.9 $\pm$ 0.1 & 1.00                   \\
\multicolumn{7}{c}{\ion{Si}{2}i $\lambda$1207 ($W_{obs}$ = 4.68 \AA)}                                              \\
 1............        & 4194.8 & 2.47687 & 5567 & 15.14 $\pm$ 0.30 & 20.1 $\pm$ 1.4 & 1.00                   \\
 2............        & 4195.8 & 2.47770 & 5496 & 13.97 $\pm$ 0.01 & 84.7 $\pm$ 1.6 & 1.00                   \\
 3............        & 4196.9 & 2.47857 & 5420 & 14.31 $\pm$ 2.23 &  6.9 $\pm$ 4.5 & 1.00                   \\
 4............        & 4197.4 & 2.47903 & 5381 & 12.57 $\pm$ 0.04 &  8.4 $\pm$ 0.8 & 1.00                   \\
 5............        & 4198.0 & 2.47949 & 5341 & 13.71 $\pm$ 1.57 &  4.6 $\pm$ 2.7 & 1.00                   \\
 6............        & 4198.3 & 2.47976 & 5318 & 14.43 $\pm$ 0.57 &  3.1 $\pm$ 0.7 & 1.00                   \\
 7............        & 4198.3 & 2.47971 & 5322 & 13.06 $\pm$ 0.03 & 32.7 $\pm$ 1.4 & 1.00                   \\
 8............        & 4199.2 & 2.48052 & 5252 & 13.99 $\pm$ 0.38 &  5.6 $\pm$ 0.7 & 1.00                   \\
 9............        & 4200.1 & 2.48122 & 5192 & 13.10 $\pm$ 2.41 &  2.2 $\pm$ 3.2 & 1.00                   \\
 10...........        & 4200.3 & 2.48136 & 5180 & 12.51 $\pm$ 0.04 &  4.8 $\pm$ 0.7 & 1.00                   \\
\multicolumn{7}{c}{\ion{C}{2} $\lambda$1335($W_{obs}$ = 2.07 \AA)}                                                 \\
 1............        & 4639.4 & 2.47642 & 5606 & 13.52 $\pm$ 0.07 &  2.6 $\pm$ 0.3 & 1.00                   \\
 2............        & 4641.7 & 2.47811 & 5460 & 14.06 $\pm$ 0.01 & 11.9 $\pm$ 0.2 & 1.00                   \\
 3............        & 4642.2 & 2.47852 & 5425 & 14.41 $\pm$ 0.03 & 13.9 $\pm$ 0.4 & 1.00                   \\
 4............        & 4642.5 & 2.47871 & 5408 & 14.15 $\pm$ 2.21 &  2.6 $\pm$ 3.4 & 1.00                   \\
 5............        & 4642.9 & 2.47904 & 5380 & 12.98 $\pm$ 0.02 &  9.9 $\pm$ 0.9 & 1.00                   \\
 6............        & 4643.4 & 2.47944 & 5345 & 13.08 $\pm$ 0.02 &  7.0 $\pm$ 0.6 & 1.00                   \\
 7............        & 4643.8 & 2.47970 & 5323 & 13.97 $\pm$ 0.08 &  4.3 $\pm$ 0.9 & 1.00                   \\
 8............        & 4643.9 & 2.47981 & 5314 & 12.72 $\pm$ 1.02 & 10.6 $\pm$13.0 & 1.00                   \\
 9............        & 4644.7 & 2.48037 & 5265 & 12.26 $\pm$ 0.70 &  1.1 $\pm$ 5.1 & 1.00                   \\
 10...........        & 4644.9 & 2.48052 & 5252 & 14.92 $\pm$ 0.80 &  3.6 $\pm$ 0.9 & 1.00                   \\
 11...........        & 4645.8 & 2.48120 & 5194 & 15.33 $\pm$ 1.70 &  2.7 $\pm$ 1.2 & 1.00                   \\
 12...........        & 4646.0 & 2.48134 & 5182 & 13.89 $\pm$ 0.08 &  3.7 $\pm$ 0.6 & 1.00                   \\
 13...........        & 4646.3 & 2.48162 & 5158 & 12.72 $\pm$ 0.03 &  5.0 $\pm$ 0.8 & 1.00                   \\
\multicolumn{7}{c}{\ion{Al}{2} $\lambda$1671 ($W_{obs}$ = 0.90 \AA)}                                               \\
 1............        & 5811.2 & 2.47814 & 5458 & 11.95 $\pm$ 0.01 & 10.6 $\pm$ 0.4 & 1.00                   \\
 2............        & 5811.9 & 2.47856 & 5421 & 12.92 $\pm$ 0.02 &  8.8 $\pm$ 0.1 & 1.00                   \\
 3............        & 5813.4 & 2.47944 & 5345 & 11.28 $\pm$ 0.03 &  4.2 $\pm$ 0.9 & 1.00                   \\
 4............        & 5813.9 & 2.47972 & 5321 & 11.70 $\pm$ 0.01 &  4.0 $\pm$ 0.4 & 1.00                   \\
 5............        & 5815.2 & 2.48053 & 5252 & 12.57 $\pm$ 3.50 &  1.5 $\pm$ 3.2 & 1.00                   \\
 6............        & 5816.4 & 2.48121 & 5193 & 11.84 $\pm$ 0.18 &  1.7 $\pm$ 1.1 & 1.00                   \\
 7............        & 5816.6 & 2.48135 & 5181 & 11.50 $\pm$ 0.04 &  3.5 $\pm$ 1.2 & 1.00                   \\
\multicolumn{7}{c}{\ion{Si}{2} $\lambda$1260 ($W_{obs}$ = 1.85 \AA)}                                               \\
 1............        & 4381.3 & 2.47607 & 5636 & 12.08 $\pm$ 0.02 &  4.1 $\pm$ 0.6 & 1.00                   \\
 2............        & 4381.7 & 2.47641 & 5607 & 13.26 $\pm$ 1.78 &  1.8 $\pm$ 1.5 & 1.00                   \\
 3............        & 4383.6 & 2.47791 & 5477 & 12.55 $\pm$ 2.19 &  1.5 $\pm$ 3.3 & 1.00                   \\
 4............        & 4383.9 & 2.47813 & 5458 & 13.11 $\pm$ 0.01 &  9.0 $\pm$ 0.4 & 1.00                   \\
 5............        & 4384.3 & 2.47845 & 5431 & 15.74 $\pm$ 0.39 &  1.9 $\pm$ 0.1 & 1.00                   \\
 6............        & 4384.6 & 2.47865 & 5414 & 13.48 $\pm$ 0.37 &  5.8 $\pm$ 1.9 & 1.00                   \\
 7............        & 4384.8 & 2.47887 & 5395 & 12.07 $\pm$ 0.04 &  3.2 $\pm$ 1.5 & 1.00                   \\
 8............        & 4385.1 & 2.47905 & 5379 & 11.97 $\pm$ 0.03 &  3.7 $\pm$ 0.9 & 1.00                   \\
 9............        & 4385.6 & 2.47944 & 5345 & 12.22 $\pm$ 0.02 &  5.9 $\pm$ 0.5 & 1.00                   \\
 10...........        & 4385.9 & 2.47971 & 5322 & 13.03 $\pm$ 0.07 &  4.1 $\pm$ 0.3 & 1.00                   \\
 11...........        & 4386.9 & 2.48052 & 5252 & 14.13 $\pm$ 0.28 &  3.1 $\pm$ 0.3 & 1.00                   \\
 12...........        & 4387.8 & 2.48122 & 5192 & 13.63 $\pm$ 1.30 &  3.2 $\pm$ 2.0 & 1.00                   \\
 13...........        & 4388.0 & 2.48135 & 5181 & 12.93 $\pm$ 4.77 &  2.0 $\pm$ 6.3 & 1.00                   \\
\multicolumn{7}{c}{\ion{Fe}{2} $\lambda$1608 ($W_{obs}$ = 0.13 \AA)}                                               \\
 1............        & 5595.1 & 2.47857 & 5420 & 13.65 $\pm$ 0.01 &  4.6 $\pm$ 0.1 & 1.00                   \\
\multicolumn{7}{c}{\ion{O}{1} $\lambda$1302 ($W_{obs}$ = 0.29 \AA)}                                                 \\
 1............        & 4529.7 & 2.47855 & 5422 & 14.58 $\pm$ 0.03 &  6.9 $\pm$ 0.2 & 1.00                   \\
 2............        & 4533.2 & 2.48125 & 5190 & 13.41 $\pm$ 0.02 & 10.7 $\pm$ 0.8 & 1.00                   \\
\multicolumn{7}{c}{\ion{Si}{2} $\lambda$1527 ($W_{obs}$ = 2.68 \AA)$^a$}                                           \\
 1............        & 5303.3 & 2.47369 & 5841 & 13.05            & 21.0           & 1.00                   \\
 2............        & 5304.0 & 2.47416 & 5801 & 13.12            & 22.8           & 1.00                   \\
 3............        & 5304.9 & 2.47476 & 5749 & 12.95            & 21.4           & 1.00                   \\
 4............        & 5306.4 & 2.47570 & 5668 & 13.62            & 11.1           & 1.00                   \\
 5............        & 5307.6 & 2.47647 & 5602 & 13.04            & 13.1           & 1.00                   \\
 6............        & 5310.1 & 2.47815 & 5457 & 13.24            & 14.2           & 1.00                   \\
 7............        & 5310.8 & 2.47857 & 5420 & 14.07            & 11.2           & 1.00                   \\
 8............        & 5312.6 & 2.47976 & 5318 & 13.17            & 12.3           & 1.00                   \\
 9............        & 5313.8 & 2.48055 & 5250 & 13.43            &  5.2           & 1.00                   \\
 10...........        & 5314.8 & 2.48124 & 5190 & 13.22            &  7.5           & 1.00                   \\
 11...........        & 5315.2 & 2.48147 & 5171 & 13.46            & 11.6           & 1.00                   \\
\\
\tableline		      			      
\multicolumn{7}{c}{System C : $z_{abs}=2.5370$}
\\
\tableline
\multicolumn{7}{c}{\ion{C}{4} $\lambda$1548 / \ion{C}{4} $\lambda$1551 ($W_{obs}$ = 0.24 / 0.14 \AA)}                    \\
 1............        & 5475.9 & 2.53693 &  430 & 13.41 $\pm $0.05 & 10.9 $\pm$ 0.2 & 0.94$^{+0.08}_{-0.07}$ \\
\\
\tableline		      			      
\multicolumn{7}{c}{System D : $z_{abs}=2.5532$}       
\\			      			      
\tableline
\multicolumn{7}{c}{\ion{C}{4} $\lambda$1548 / \ion{C}{4} $\lambda$1551 ($W_{obs}$ = 1.40 / 1.23 \AA)}                    \\
 1............        & 5498.0 & 2.55125 & --782 & 13.18 $\pm$ 0.22 &  3.0 $\pm$ 0.7 & 0.30$^{+0.14}_{-0.09}$ \\
 2............        & 5500.1 & 2.55261 & --897 & 13.05 $\pm$ 0.14 &  4.5 $\pm$ 0.6 & 0.81$^{+0.24}_{-0.18}$ \\
 3............        & 5500.4 & 2.55280 & --913 & 13.54 $\pm$ 0.15 &  8.2 $\pm$ 3.3 & 0.23$^{+0.05}_{-0.05}$ \\
 4............        & 5501.0 & 2.55314 & --942 & 14.26 $\pm$ 0.02 & 11.7 $\pm$ 0.5 & 0.99$^{+0.02}_{-0.02}$ \\
 5............        & 5501.5 & 2.55351 & --973 & 16.66 $\pm$ 0.06 &  5.6 $\pm$ 0.1 & 0.99$^{+0.02}_{-0.02}$ \\
\multicolumn{7}{c}{\ion{Si}{4} $\lambda$1394 / \ion{Si}{4} $\lambda$1403 ($W_{obs}$ = 0.49 / 0.34 \AA)}                  \\
 1............        & 4952.2 & 2.55313 & --941 & 13.05 $\pm$ 0.06 &  9.4 $\pm$ 0.3 & 0.66$^{+0.06}_{-0.06}$ \\
 2............        & 4952.7 & 2.55350 & --972 & 13.41 $\pm$ 0.02 &  8.4 $\pm$ 0.1 & 0.96$^{+0.03}_{-0.03}$ \\
\\			      			      
\tableline		      			      
\multicolumn{7}{c}{System E : $z_{abs}=1.8875$}
\\			      			      
\tableline		      			      
\multicolumn{7}{c}{\ion{C}{4}  $\lambda$1548 / \ion{C}{4} $\lambda$1551 ($W_{obs}$ = 0.67 / 0.41 \AA)}                   \\
 1............        & 4469.8 & 1.88708 &60493 & 13.07 $\pm$ 0.30 &  7.5 $\pm$ 1.3 & 0.73$^{+1.05}_{-0.43}$ \\
 2............        & 4470.1 & 1.88730 &60471 & 13.61 $\pm$ 0.08 & 14.2 $\pm$ 2.3 & 0.93$^{+0.08}_{-0.08}$ \\
 3............        & 4470.5 & 1.88753 &60448 & 13.35 $\pm$ 0.16 &  8.4 $\pm$ 2.7 & 0.64$^{+0.18}_{-0.16}$ \\
 4............        & 4470.7 & 1.88771 &60430 & 13.06 $\pm$ 0.33 &  7.7 $\pm$ 3.1 & 0.78$^{+1.09}_{-0.45}$ \\
 5............        & 4471.0 & 1.88788 &60414 & 13.34 $\pm$ 0.17 & 10.8 $\pm$ 1.6 & 0.67$^{+0.26}_{-0.19}$ \\
\\			      			      
\tableline		      			      
\multicolumn{7}{c}{System F : $z_{abs}=1.9644$}       
\\			      			      
\tableline		      			      
\multicolumn{7}{c}{\ion{C}{4}  $\lambda$1548 / \ion{C}{4} $\lambda$1551 ($W_{obs}$ = 1.41 / 1.08 \AA)}                     \\
 1............        & 4587.4 & 1.96305 &52980 & 13.49 $\pm$ 0.12 & 11.2 $\pm$ 0.9 & 0.52$^{+0.10}_{-0.09}$   \\
 2............        & 4587.7 & 1.96327 &52959 & 13.86 $\pm$ 0.02 &  5.9 $\pm$ 0.2 & 1.00                     \\
 3............        & 4588.0 & 1.96345 &52941 & 13.78 $\pm$ 0.01 &  7.5 $\pm$ 0.3 & 1.00                     \\
 4............        & 4588.6 & 1.96381 &52906 & 13.04 $\pm$ 0.44 &  5.1 $\pm$ 0.9 & 0.37$^{+\infty}_{-0.28}$ \\
 5............        & 4589.4 & 1.96436 &52852 & 14.00 $\pm$ 0.03 &  8.3 $\pm$ 0.2 & 0.99$^{+0.03}_{-0.03}$   \\
 6............        & 4589.8 & 1.96459 &52829 & 13.26 $\pm$ 0.17 &  9.4 $\pm$ 1.2 & 0.71$^{+0.30}_{-0.22}$   \\
 7............        & 4590.2 & 1.96486 &52803 & 13.40 $\pm$ 0.08 &  8.3 $\pm$ 0.5 & 0.89$^{+0.11}_{-0.10}$   \\
 8............        & 4590.5 & 1.96506 &52783 & 12.85 $\pm$ 0.02 &  5.9 $\pm$ 0.4 & 1.00                     \\
\\			      			      
\tableline		      			      
\multicolumn{7}{c}{System G : $z_{abs}=2.0701$}       
\\			      			      
\tableline
\multicolumn{7}{c}{\ion{C}{4}  $\lambda$1548 / \ion{C}{4} $\lambda$1551 ($W_{obs}$ = 1.24 / 0.74 \AA)}                   \\
 1............        & 4752.1 & 2.06947 &42665 & 12.70 $\pm$ 0.03 & 11.6 $\pm$ 1.1 & 1.00                   \\
 2............        & 4752.6 & 2.06974 &42639 & 13.75 $\pm$ 0.00 & 10.0 $\pm$ 0.1 & 1.00                   \\
 3............        & 4752.9 & 2.06997 &42617 & 14.04 $\pm$ 0.04 & 14.8 $\pm$ 1.4 & 1.00$^{+0.02}_{-0.02}$ \\
 4............        & 4753.2 & 2.07016 &42599 & 13.47 $\pm$ 0.01 & 12.0 $\pm$ 0.2 & 1.00                   \\
 5............        & 4753.7 & 2.07045 &42571 & 13.01 $\pm$ 0.02 &  9.1 $\pm$ 0.6 & 1.00                   \\
 6............        & 4754.0 & 2.07070 &42547 & 12.56 $\pm$ 0.05 & 11.8 $\pm$ 1.8 & 1.00                   \\
\multicolumn{7}{c}{\ion{Si}{4} $\lambda$1403 ($W_{obs}$ = 0.23 \AA)}                                               \\
 1............        & 4306.3 & 2.06984 &42629 & 13.16 $\pm$ 0.02 & 14.6 $\pm$ 0.7 & 1.00                   \\
 2............        & 4306.6 & 2.07010 &42604 & 12.83 $\pm$ 0.03 &  7.3 $\pm$ 0.6 & 1.00                   \\
\\
\tableline		      			      
\multicolumn{7}{c}{System H : $z_{abs}=2.2653$}       
\\			      			      
\tableline
\multicolumn{7}{c}{\ion{C}{4} $\lambda$1548 / \ion{C}{4} $\lambda$1551 ($W_{obs}$ = 0.14 / 0.09 \AA)}                    \\
 1............        & 5055.2 & 2.26521 &24357 & 13.05 $\pm$ 0.22 &  5.9 $\pm$ 0.5 & 0.75$^{+0.40}_{-0.28}$ \\
 2............        & 5055.5 & 2.26542 &24337 & 13.85 $\pm$ 0.16 & 13.8 $\pm$ 2.8 & 0.14$^{+0.03}_{-0.02}$ \\
\multicolumn{7}{c}{\ion{Si}{2}i $\lambda$1207 ($W_{obs}$ = 0.17 \AA)}                                              \\
 1............        & 3939.5 & 2.26520 &24357 & 12.76 $\pm$ 0.13 &  4.2 $\pm$ 0.8 & 1.00                   \\
 2............        & 3939.6 & 2.26534 &24345 & 11.82 $\pm$ 0.21 &  1.7 $\pm$ 3.6 & 1.00                   \\
\\
\tableline		      			      
\multicolumn{7}{c}{System I : $z_{abs}=2.1753$}       
\\			      			      
\tableline
\multicolumn{7}{c}{\ion{C}{4}  $\lambda$1548 / \ion{C}{4} $\lambda$1551 ($W_{obs}$ = 0.66 / 0.40 \AA)}                    \\
 1............        & 4914.9 & 2.17460 & 32722 & 12.83 $\pm$ 0.01 &  9.9 $\pm$ 0.5 & 1.00                   \\
 2............        & 4915.2 & 2.17481 & 32703 & 12.97 $\pm$ 0.22 &  8.7 $\pm$ 1.4 & 0.90$^{+0.84}_{-0.43}$ \\
 3............        & 4915.4 & 2.17494 & 32691 & 12.25 $\pm$ 0.03 &  2.0 $\pm$ 1.3 & 1.00                   \\
 4............        & 4915.8 & 2.17520 & 32666 & 13.33 $\pm$ 0.11 & 14.8 $\pm$ 0.7 & 0.70$^{+0.24}_{-0.16}$ \\
 5............        & 4916.2 & 2.17542 & 32646 & 13.17 $\pm$ 0.46 &  0.8 $\pm$ 0.6 & 0.37$^{+1.04}_{-0.26}$ \\
 6............        & 4916.4 & 2.17558 & 32631 & 13.65 $\pm$ 0.02 &  8.6 $\pm$ 0.3 & 0.87$^{+0.02}_{-0.02}$ \\
 7............        & 4916.7 & 2.17578 & 32612 & 12.76 $\pm$ 0.03 & 27.4 $\pm$ 2.4 & 1.00                   \\
\multicolumn{7}{c}{\ion{Si}{4} $\lambda$ 1394 / \ion{Si}{4} $\lambda$1403 ($W_{obs}$ = 0.18 / 0.07 \AA)}                  \\
 1............        & 4424.6 & 2.17462 & 32720 & 11.96 $\pm$ 0.05 &  8.2 $\pm$ 1.7 & 1.00                   \\
 2............        & 4424.9 & 2.17480 & 32704 & 11.67 $\pm$ 0.08 &  1.2 $\pm$ 4.9 & 1.00                   \\
 3............        & 4425.1 & 2.17495 & 32690 & 11.74 $\pm$ 0.08 &  4.7 $\pm$ 2.1 & 1.00                   \\
 4............        & 4425.4 & 2.17519 & 32667 & 12.04 $\pm$ 0.04 &  9.3 $\pm$ 1.2 & 1.00                   \\
 5............        & 4426.0 & 2.17557 & 32632 & 12.51 $\pm$ 0.01 &  7.2 $\pm$ 0.3 & 1.00                   \\
\multicolumn{7}{c}{\ion{Si}{2}i $\lambda$1207 ($W_{obs}$ = 0.27 \AA)}                                               \\
 1............        & 3830.9 & 2.17524 & 32663 & 12.35 $\pm$ 0.03 &  7.7 $\pm$ 0.8 & 1.00                   \\
 2............        & 3831.4 & 2.17560 & 32629 & 12.61 $\pm$ 0.03 &  6.8 $\pm$ 0.5 & 1.00                   \\
\multicolumn{7}{c}{\ion{Si}{2} $\lambda$1527}                                                                       \\
 1............        &\multicolumn{5}{c}{(blending with \ion{Si}{4} $\lambda$1394 of System B)} &                 \\
\enddata
\tablenotetext{a}{Parameters are obtained by automatic fitting without
                  considering \ion{C}{4} mini-BAL components.  Thus,
                  equivalent width is the upper limit.}
\end{deluxetable}
\clearpage

\begin{figure}
 \begin{center}
  \includegraphics[width=17cm,angle=0]{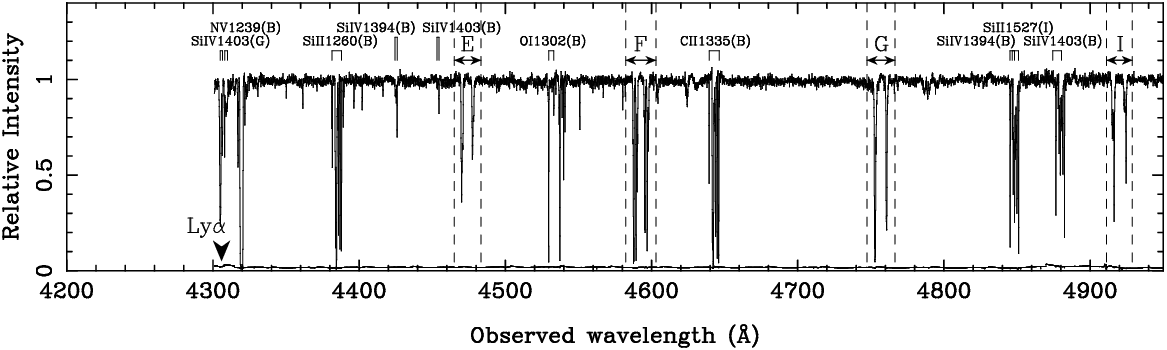}
  \includegraphics[width=17cm,angle=0]{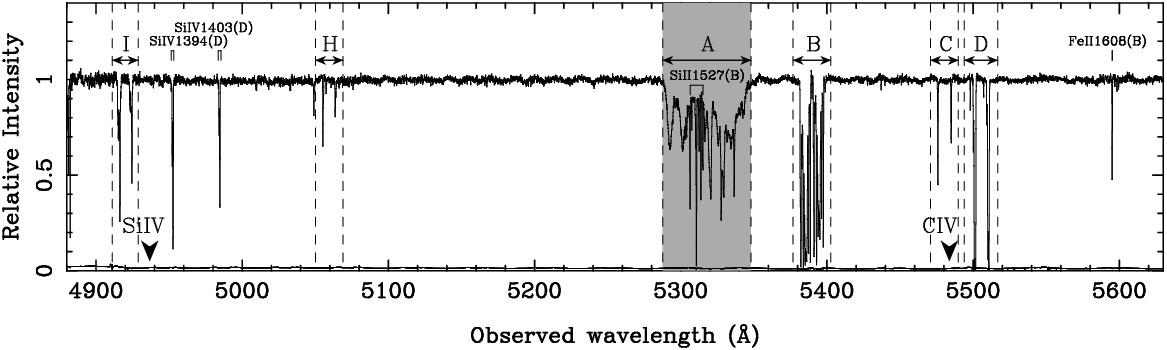}
  \includegraphics[width=17cm,angle=0]{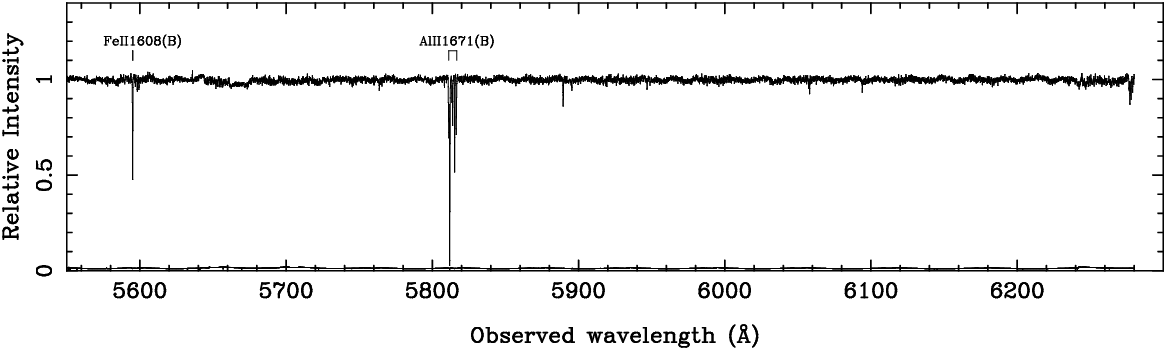}
 \end{center}
\caption{Normalized spectrum of HS~1603+3820 after rebinning to
  0.03~\AA\ per pixel, and combining all 6 spectra taken with
  Subaru+HDS (see Table~1).  \ion{C}{4} absorption systems and other
  metal lines are marked. Positions of the quasar emission lines of
  \lya, \ion{Si}{4}, and \ion{C}{4} are marked with downward
  arrows. The region blueward of the \lya\ emission line is not shown
  because the spectrum normalization by continuum fitting is not
  reliable in the \lya\ forest. The lower line is the 1$\sigma$ error
  spectrum. The region of the \ion{C}{4} mini-BAL in system~A (5288 --
  5348 \AA) is shaded.}
\end{figure}
\clearpage

\begin{figure}
 \begin{center}
  \includegraphics[width=17cm,angle=0]{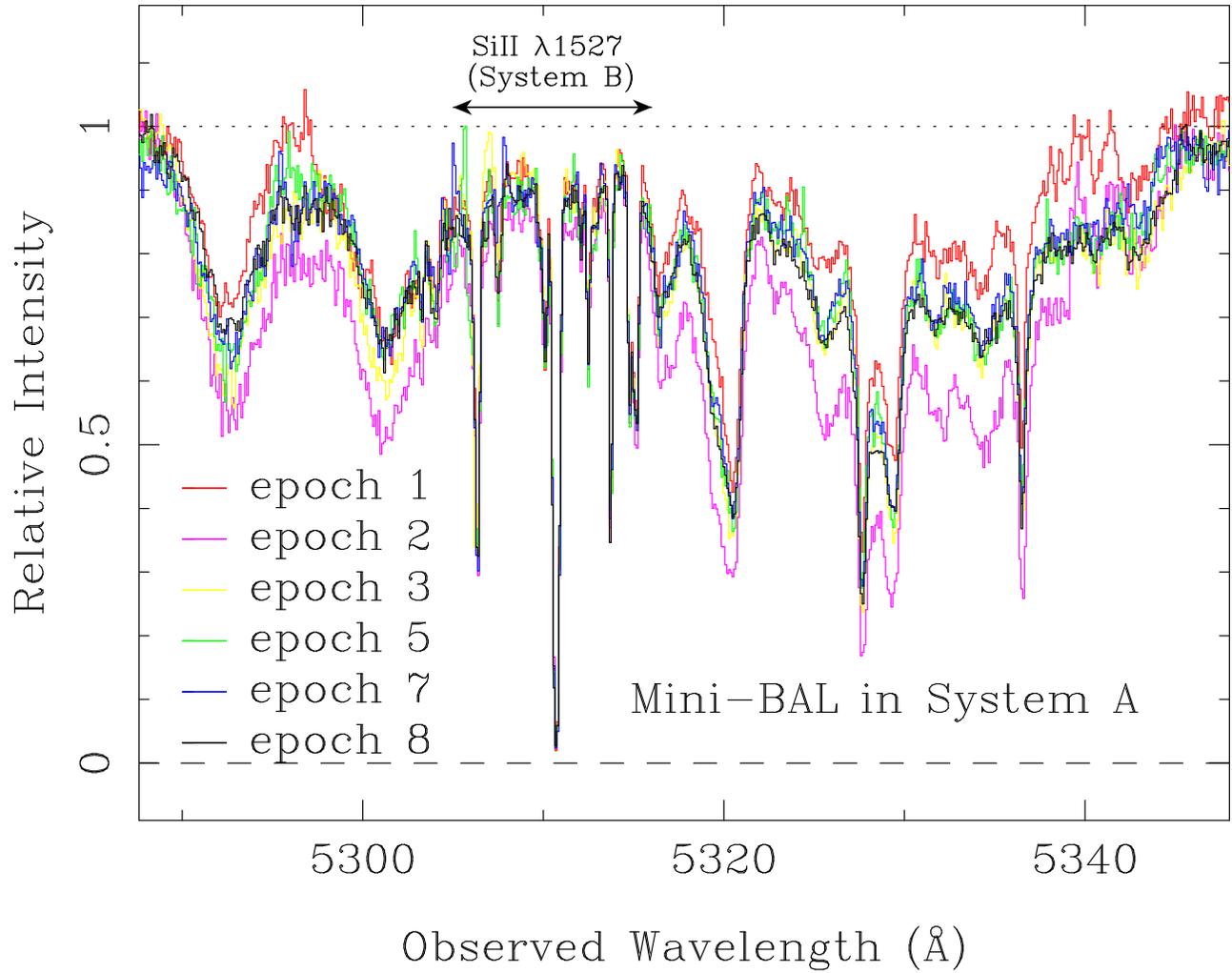}
 \end{center}
\caption{Normalized spectrum around the \ion{C}{4} mini-BAL in
  system~A, for each Subaru+HDS observation (see electronic version in
  color). The absorption profile (both the broad and narrow \ion{C}{4}
  components) have obviously changed within 1 year in the quasar
  rest-frame. On the other hand, the \ion{Si}{2} $\lambda$1527 line of
  System~B did not show significant variability.}
\end{figure}
\clearpage

\begin{figure}
 \begin{center}  
   \includegraphics[width=12cm,angle=0]{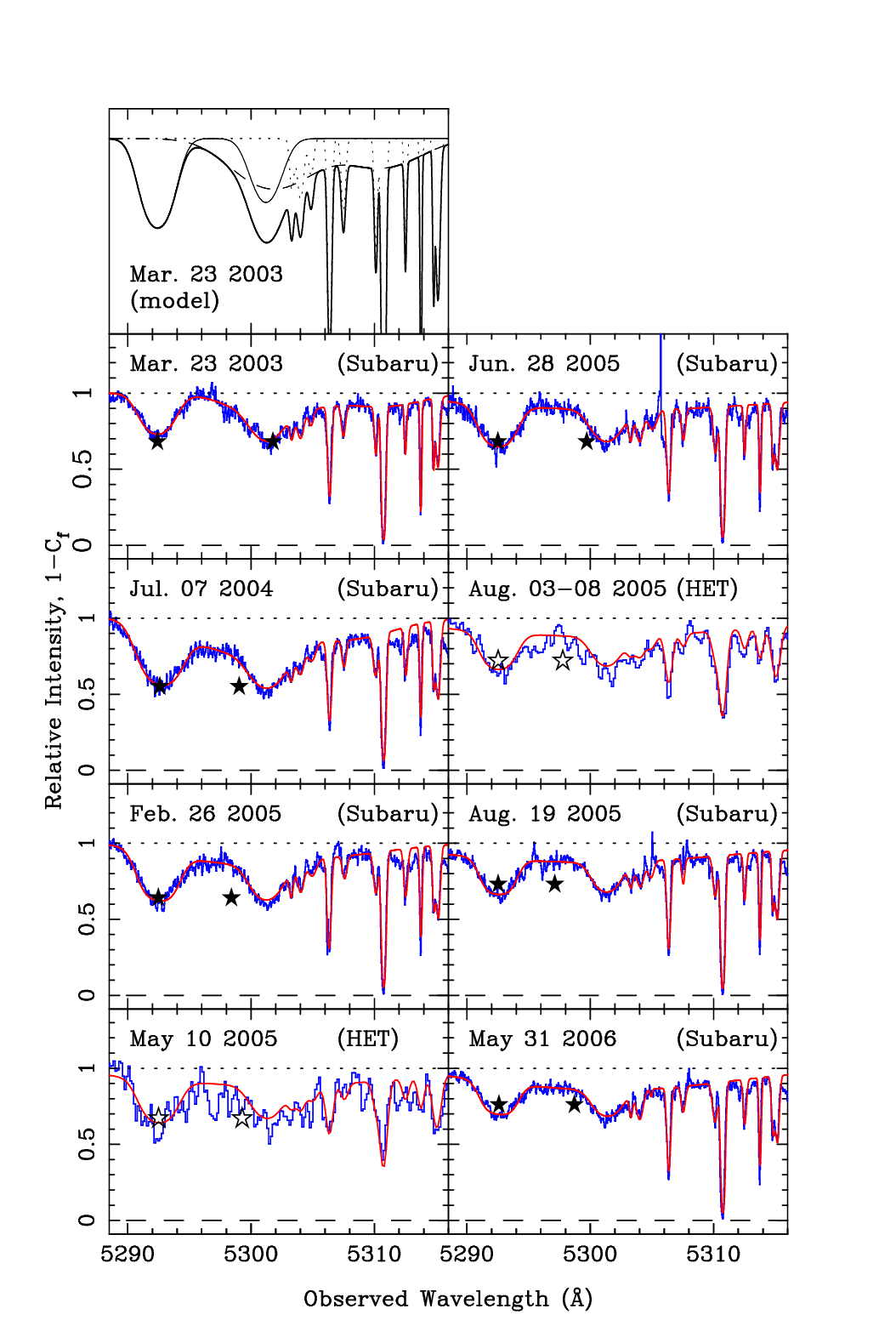}
 \end{center}
\caption{Observed spectra (blue histogram: see electronic version) and
  best-fit models (red line) of the region around \ion{C}{4} mini-BAL
  at $\zabs \sim 2.43$, for 8 observing epochs from March 23, 2002 to
  May 31, 2006. Filled stars with 1$\sigma$ errors denote the coverage
  fractions of narrow (left) and broad (right) components. The stars
  are placed at the wavelengths of the blue members of the doublet
  (i.e., \ion{C}{4} 1548). Open stars plotted in the HET spectra are
  coverage fractions that are interpolated from the Subaru spectra. In
  the top left panel, we also present models of the narrow and broad
  \ion{C}{4} components as well as the \ion{Si}{2}~$\lambda$1527
  components in System~B from the first observation.}
\end{figure}
\clearpage

\begin{figure}
\vspace{-0.8cm}
 \begin{center}
  \includegraphics[width=9.5cm,angle=0]{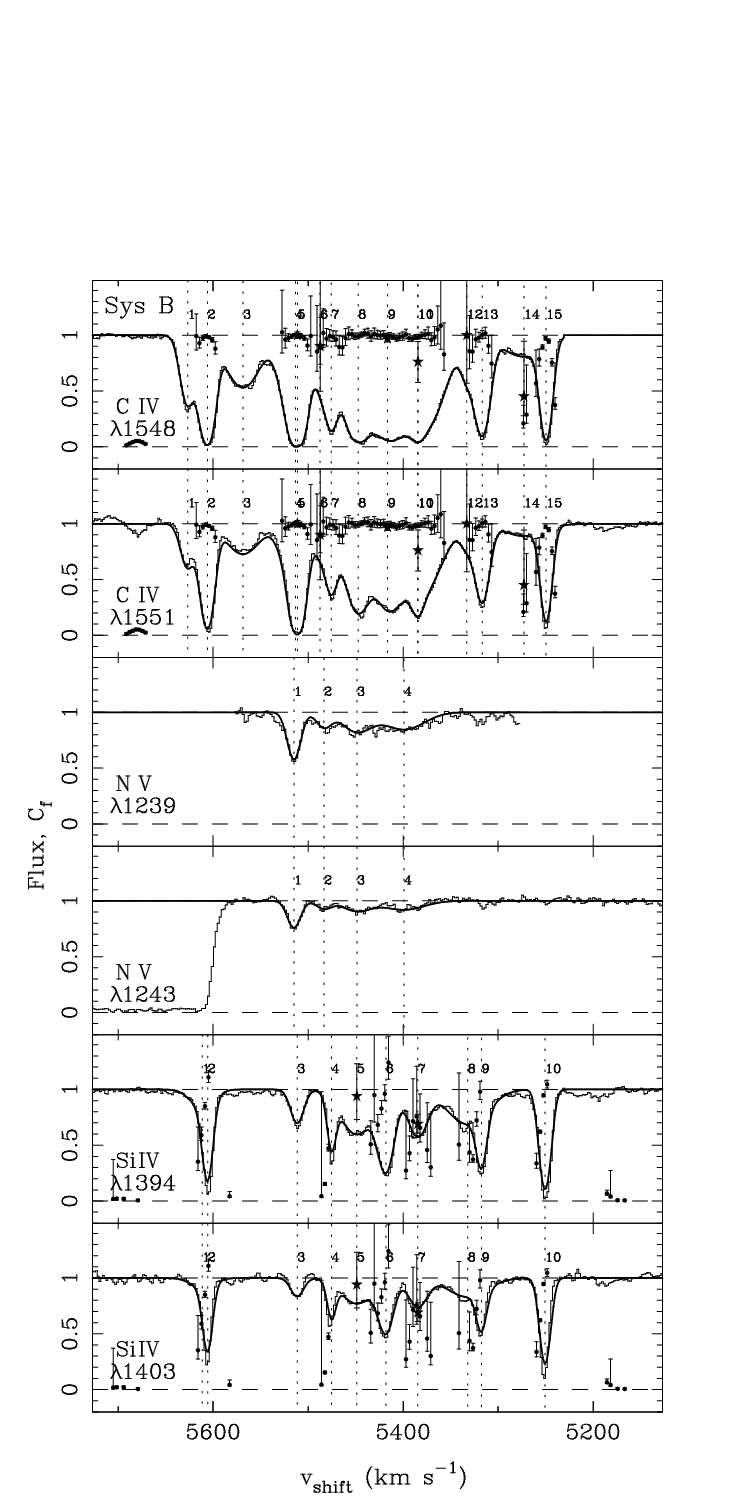}
 \end{center}
\caption{Velocity plots of various transitions for the 8 \ion{C}{4}
  NALs (systems B--I). Vertical dotted lines with ID numbers are the
  positions of absorption lines. Smooth lines are model fits using
  {\sc minfit}. Filled stars with 1$\sigma$ error bars are coverage
  fractions determined by {\sc minfit}. If a star is not displayed,
  the value of the coverage fraction is 1.0. Small dots with error
  bars denote the coverage fractions evaluated by the pixel-by-pixel
  method. The error values contain both Poisson noise and error from
  continuum level uncertainty.}
\end{figure}
\clearpage
 \begin{center}
  \includegraphics[width=9.5cm,angle=0]{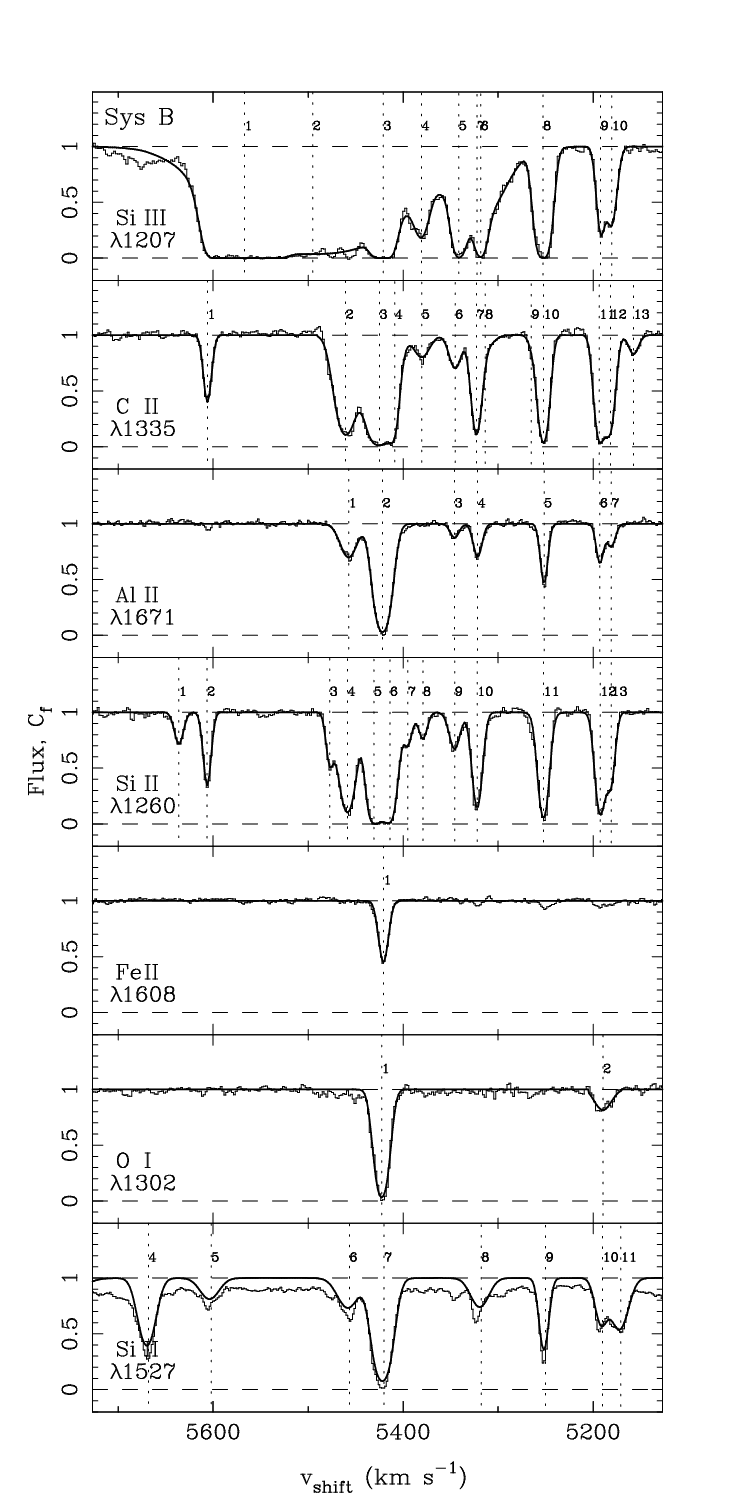}
 \\Fig.~4 --- Continued.
\clearpage
  \includegraphics[width=9.5cm,angle=0]{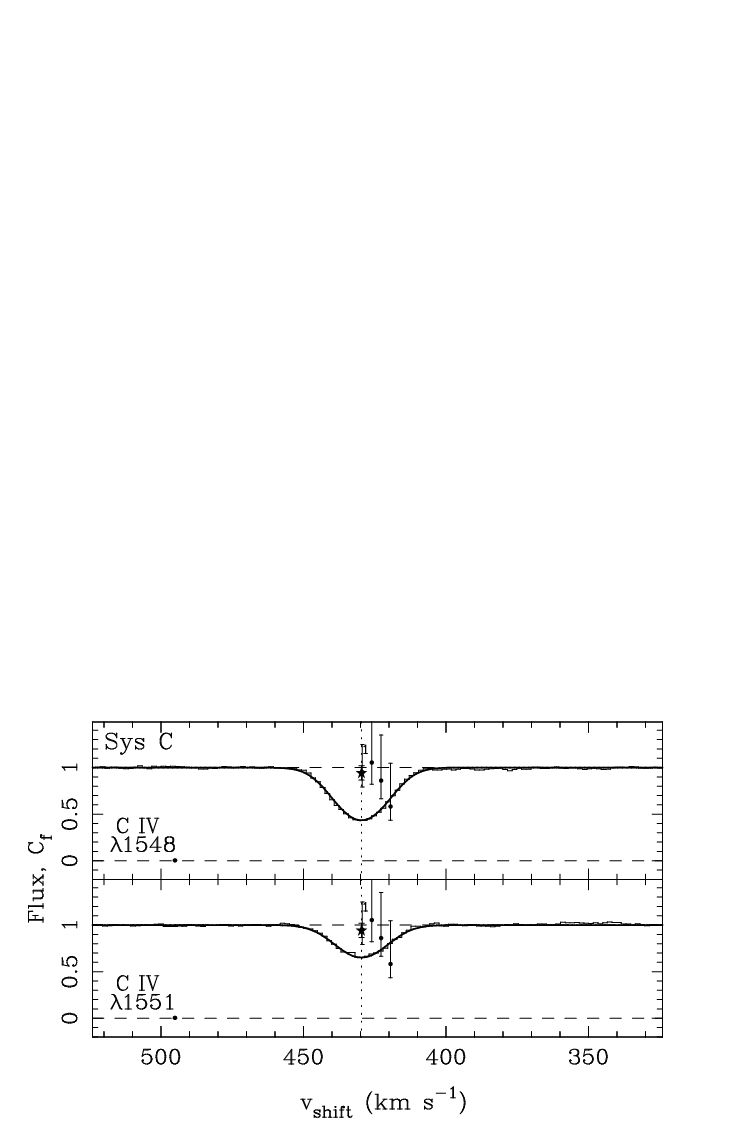}
 \\Fig.~4 --- Continued.
\clearpage
  \includegraphics[width=9.5cm,angle=0]{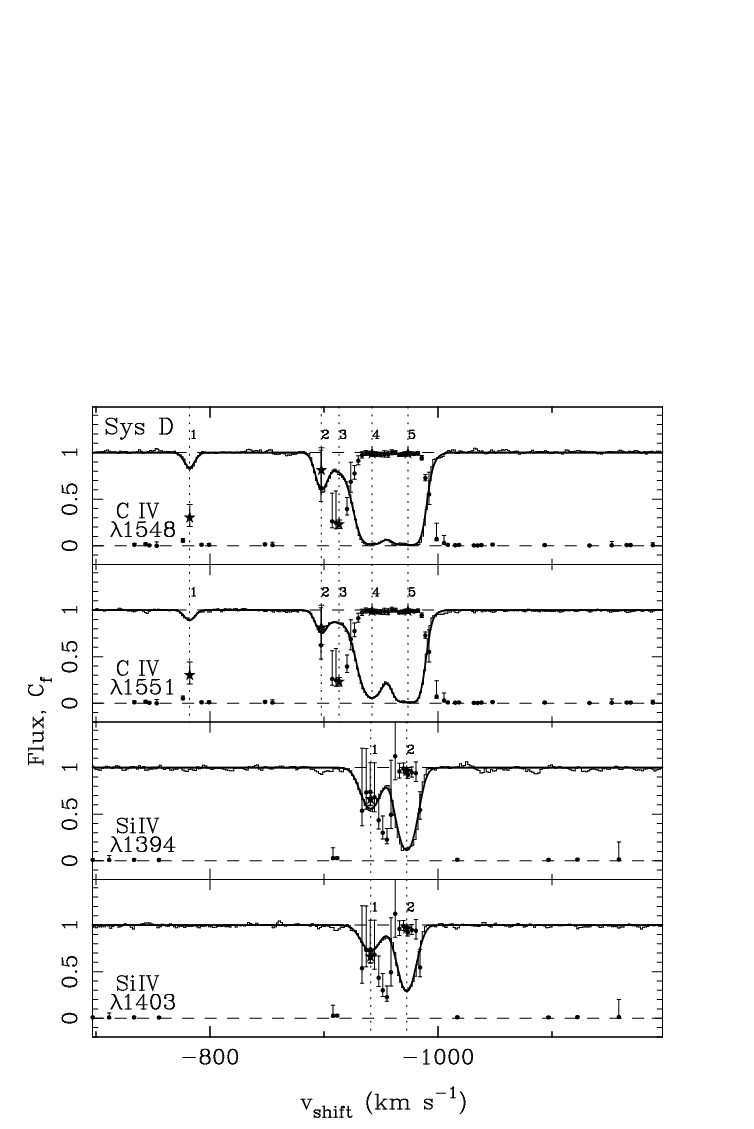}
 \\Fig.~4 --- Continued.
\clearpage
  \includegraphics[width=9.5cm,angle=0]{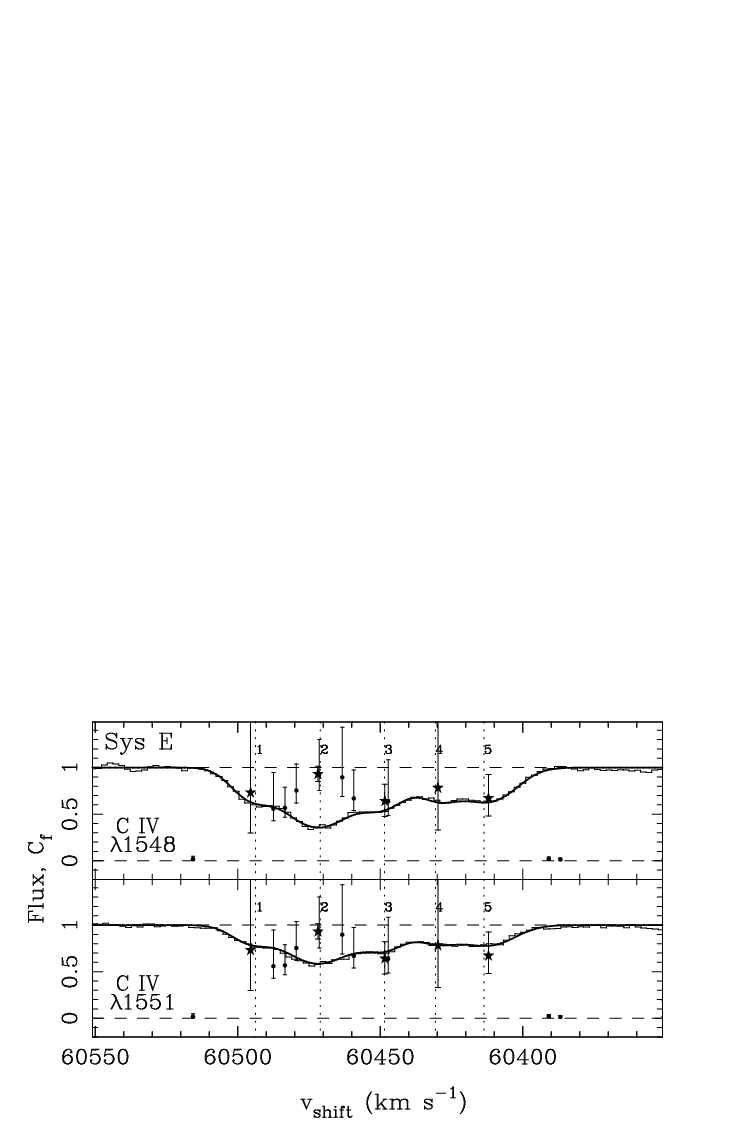}
 \\Fig.~4 --- Continued.
\clearpage
  \includegraphics[width=9.5cm,angle=0]{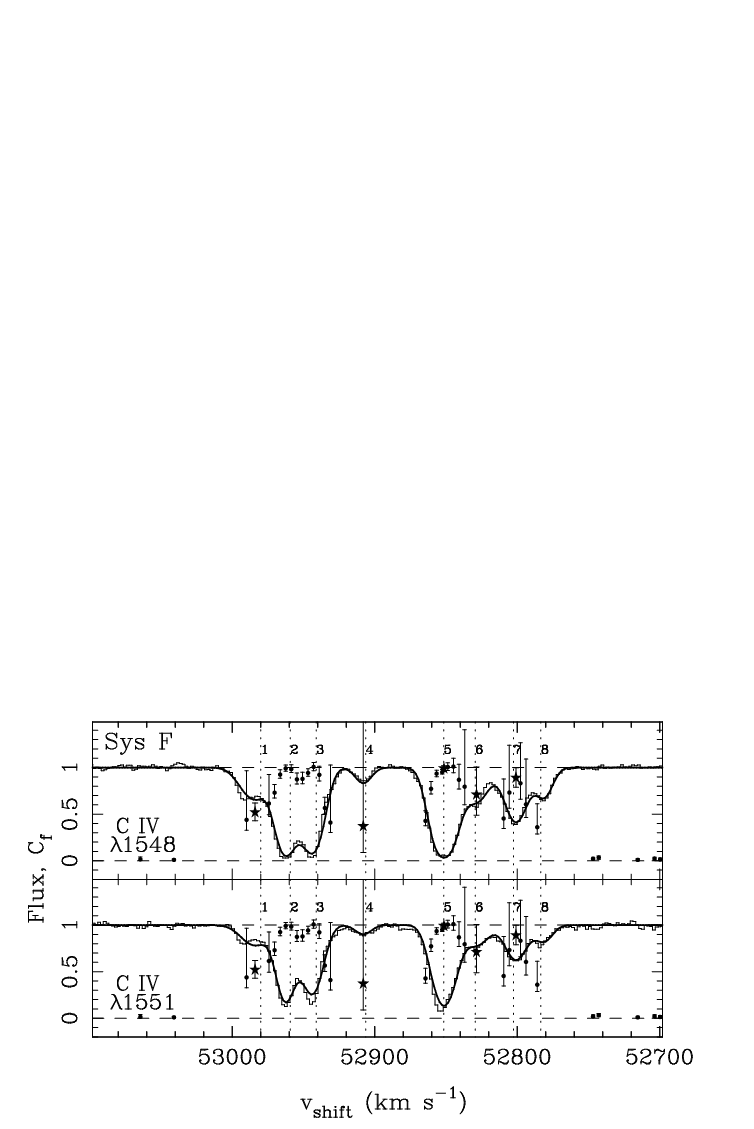}
 \\Fig.~4 --- Continued.
\clearpage
  \includegraphics[width=9.5cm,angle=0]{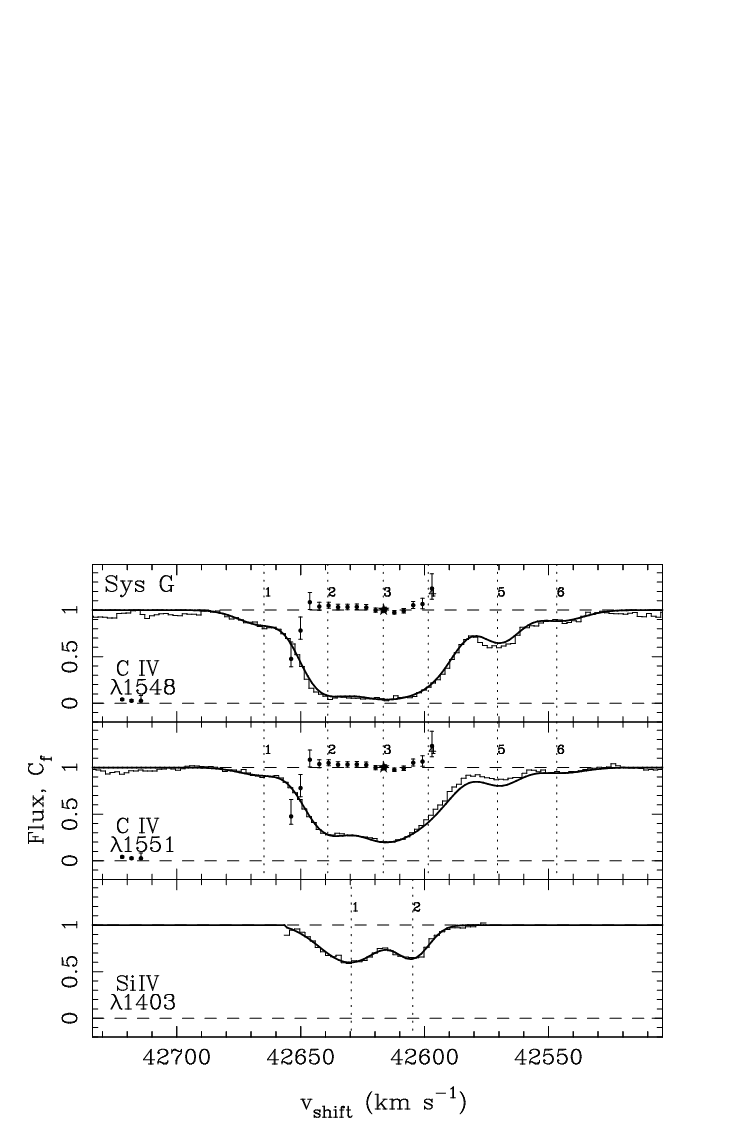}
 \\Fig.~4 --- Continued.
\clearpage
  \includegraphics[width=9.5cm,angle=0]{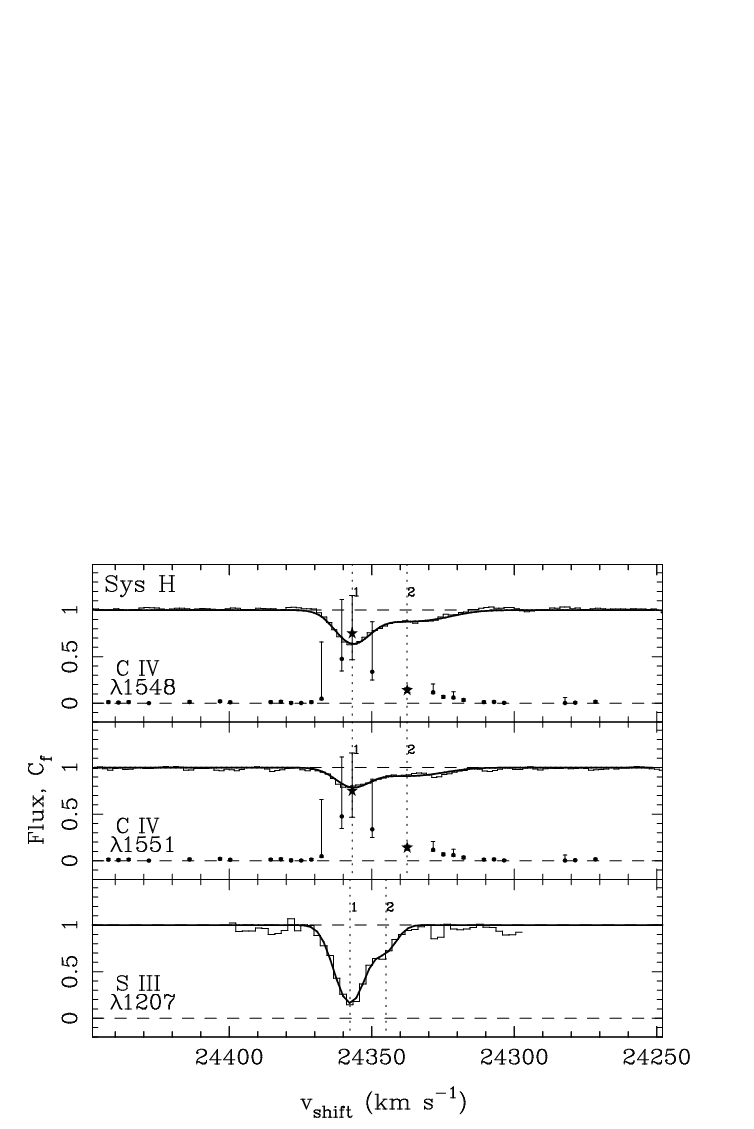}
 \\Fig.~4 --- Continued.
\clearpage
  \includegraphics[width=9.5cm,angle=0]{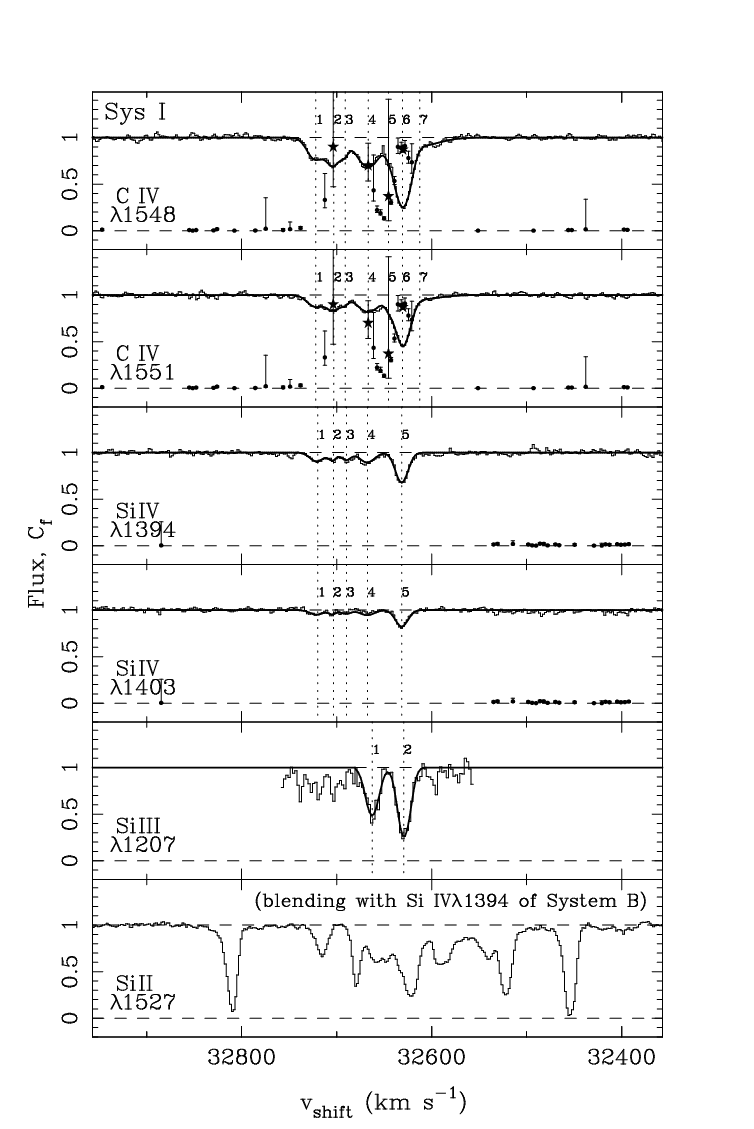}
 \\Fig.~4 --- Continued.
\clearpage
 \end{center}

\begin{figure}
 \begin{center}
  \includegraphics[width=7.5cm,angle=0]{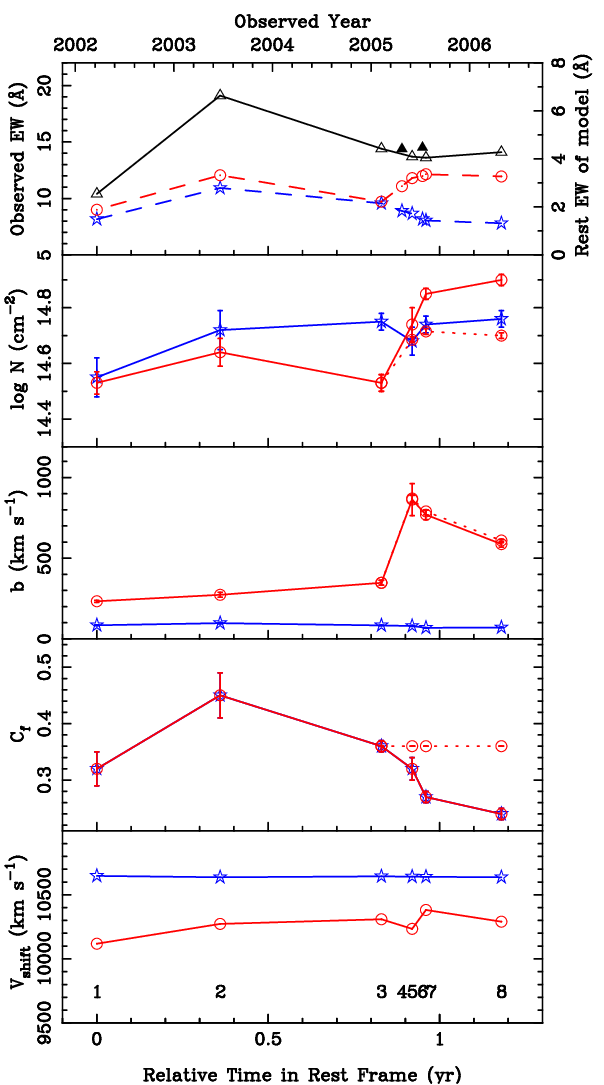}
 \end{center}
\caption{Variability of the \ion{C}{4} mini-BAL ($\zabs \sim 2.43$)
  parameters over our monitoring period. From top to bottom, the
  frames show (1) rest frame equivalent width, (2) column densities,
  (3) Doppler parameters, (4) coverage fractions, and (5) shift
  velocities. The solid line and triangles in the top panel refer to
  the equivalent width of the entire mini-BAL profile (see Figures 1
  and 2), while the model parameters refer to the bluest troughs (see
  Figure 3). The open and filled triangles in the top panel are
  measurements of the rest frame equivalent width, directly from the
  Subaru and HET data respectively.  These include all \ion{C}{4} and
  \ion{Si}{2}~$\lambda$1527 components (the \ion{Si}{2} components
  make only a small contribution). In all panels, the blue line and
  open stars refer to the narrow model component while the red line
  and open circles refer to the broad model component.  The horizontal
  axis gives the observation time, both as the observed year (top
  label) and as the time in the quasar rest-frame, relative to the
  first observation (bottom label). Only the results from Subaru
  spectra are shown, except for the equivalent width panel, because
  the low-resolution and low-S/N ratio of the HET+MRS spectra
  prevented us from fitting models. The dotted lines in the 2nd, 3rd,
  and 4th panels show the best fit parameters for epoch 3 through
  epoch 8, assuming a constant \cf\ (=0.36) for the broad component.}
\end{figure}
\clearpage

\begin{figure}
\hspace{-1cm}
 \begin{center}
  \includegraphics[width=18cm,angle=0]{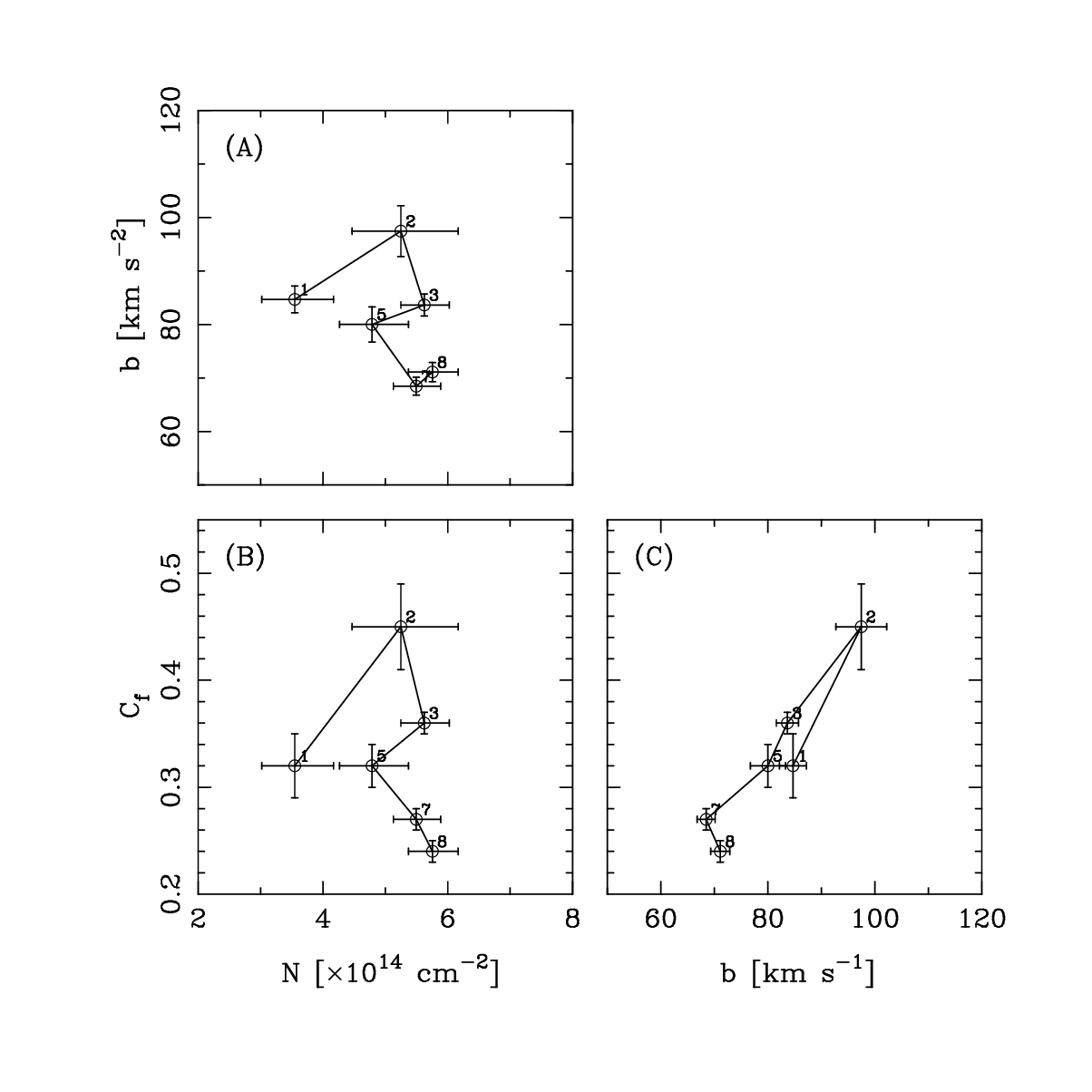}
 \end{center}
\caption{The relationship between (a) column density and Doppler
  parameter, (b) column density and coverage fraction, and (c) Doppler
  parameter and coverage fraction, with 1$\sigma$ error bars for the
  {\it narrow component} of the \ion{C}{4} mini-BAL in System~A. Each
  point is labeled with the epoch number.}
\end{figure}
\clearpage

\begin{figure}
\hspace{-1cm}
 \begin{center}
  \includegraphics[width=18cm,angle=0]{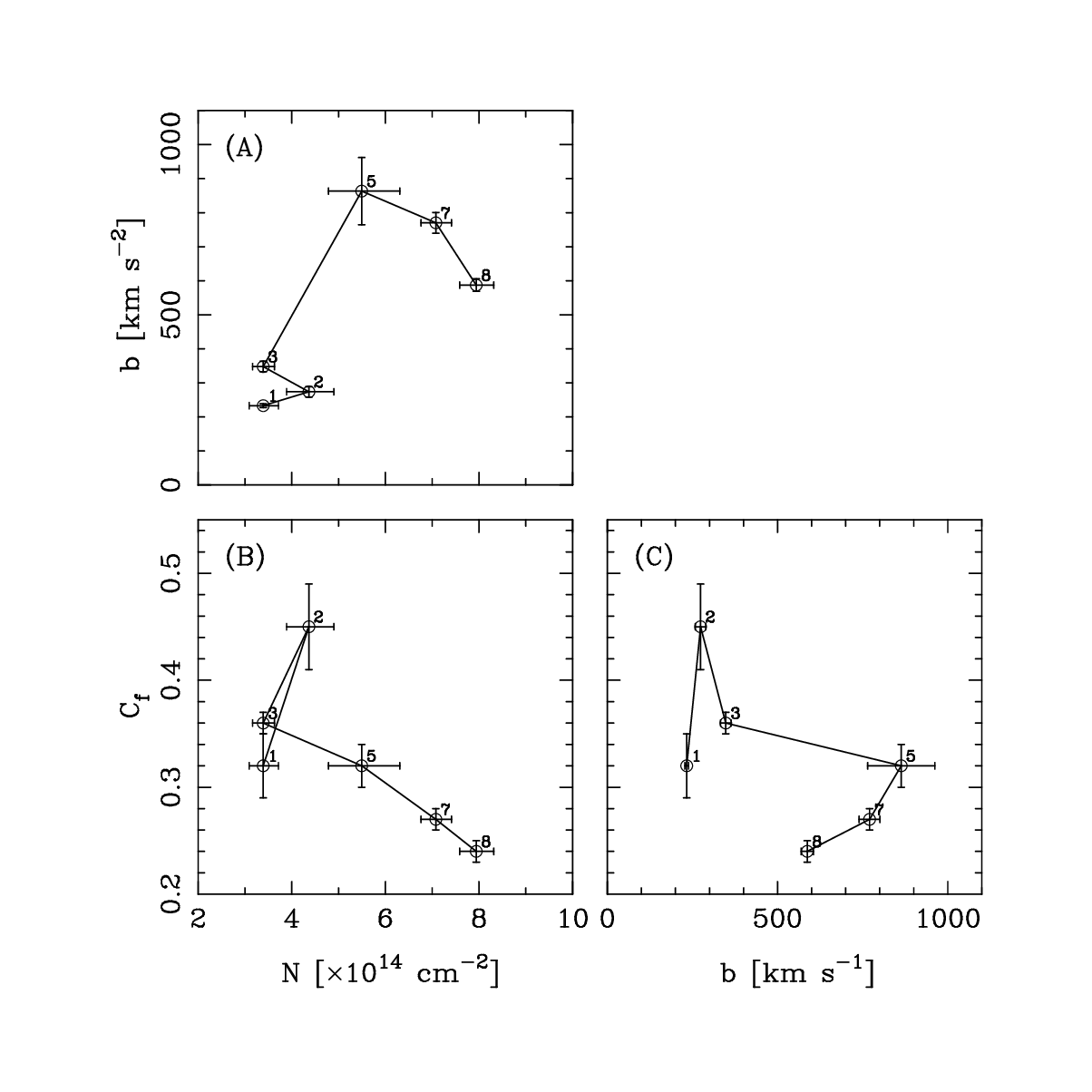}
 \end{center}
\caption{Same as Figure~6, but for the broad component of the
  \ion{C}{4} mini-BAL in System~A. The error bars here are comparable
  to those of the narrow component shown in Figure~6, although the
  line center in the broad component is highly uncertain. This is
  because the fit parameters plotted here are not strongly affected by
  the error of the line center; the column density and coverage
  fraction are mainly determined by the depth around the line center,
  and the Doppler parameter is set primarily by the line width.}
\end{figure}
\clearpage

\end{document}